%% file: paper.tex
\documentclass[conference]{IEEEtran}

\usepackage[utf8]{inputenc} % allow utf-8 input
\usepackage[T1]{fontenc}    % use 8-bit T1 fonts
\usepackage{hyperref}       % hyperlinks
\usepackage{url}            % simple URL typesetting
\usepackage{booktabs}       % professional-quality tables
\usepackage{amsfonts}       % blackboard math symbols
\usepackage{nicefrac}       % compact symbols for 1/2, etc.
\usepackage{microtype}      % microtypography
\usepackage{graphicx}
\usepackage{doi}

\usepackage{algpseudocode}
\usepackage{algorithm}
\usepackage{cite}
\usepackage{amsmath,amssymb,amsfonts,amsthm}
\usepackage{graphicx}
\usepackage{textcomp}
\usepackage{soul,xcolor}
\usepackage{caption}
\usepackage{subcaption}
\usepackage{enumitem}
\usepackage{pifont}
\usepackage{makecell}
\usepackage{multirow}
\usepackage{amsfonts}
\usepackage{titlesec}

\usepackage{cleveref}       % smart cross-referencing

\newcommand{\name}{\textsf{Rialto}}
\newcommand{\malicious}{\textsf{Rialto+}}
\newcommand{\SortingMPC}{$\mathsf{SortingMPC}$}
\newcommand{\SettlementMPC}{$\mathsf{SettlementMPC}$}
\newcommand{\ShuffleMPC}{$\mathsf{ShuffleMPC}$}

\theoremstyle{definition}
\newtheorem{definition}{Definition}[section]
\newtheorem{claim}{Claim}

\newcommand{\Sim}{\mathsf{Sim}}
\newcommand{\Chal}{\mathsf{C}}
\newcommand{\Adv}{\mathsf{A}}
\newcommand{\negl}{\mathrm{negl}}
\newcommand{\secp}{\mathsf{\kappa}}

\usepackage{bbding,stackengine,scalerel,xcolor}
\stackMath
\def\BCF{\mkern-1mu\bigcirc\mkern-1mu}

\newcommand\BC[1]{\csname BC#1\endcsname}
\newcommand\HC[2]{\csname HC#1#2\endcsname}
\newcommand\moon[3]{%
  \if W#1\def\tmp{B}\else\def\tmp{W}\fi%
  \stackengine{0pt}{%
    \stackengine{0pt}{%
      \stackengine{0pt}{\BC{\tmp}}{\scalebox{#3}[1]{\BC{#1}}}{O}{c}{F}{F}{L}%
    }{\HC{#2}{#1}}{O}{#2}{F}{F}{L}%
  }{\BCF}{O}{c}{F}{F}{L}
}

\newcommand{\cmark}{\footnotesize{\moon{B}{l}{1.}}}%
\newcommand{\xmark}{\footnotesize{\moon{W}{r}{1.}}}%
\newcommand{\hmark}{\footnotesize{\moon{B}{l}{.0}}}%

\title{Privacy-Preserving Decentralized Exchange Marketplaces}

\date{}

\author{\IEEEauthorblockN{Kavya Govindarajan}
\IEEEauthorblockA{\textit{IBM Research India}\\
kavya.g@ibm.com}
\and
\IEEEauthorblockN{Dhinakaran Vinayagamurthy}
\IEEEauthorblockA{\textit{IBM Research India}\\
dvinaya1@in.ibm.com}
\and
\IEEEauthorblockN{Praveen Jayachandran}
\IEEEauthorblockA{\textit{IBM Research India}\\
praveen.j@in.ibm.com}
\and
\IEEEauthorblockN{Chester Rebeiro}
\IEEEauthorblockA{\textit{IIT Madras, India}\\
chester@cse.iitm.ac.in}
}

\begin{document}
\maketitle

\begin{abstract}
  \input{abstract}
\end{abstract}

% keywords can be removed
\begin{IEEEkeywords}
  Decentralized Exchange, Multi-Party Computation
\end{IEEEkeywords}

\section{Introduction}
\input{intro}

\section{Problem Description}
\input{problem}

\section{Solution Overview}
\input{overview}

\section{Security Properties}
\input{security}

\section{\name: A Privacy Preserving Decentralized Marketplace}
\input{protocol}

\section{Security Analysis}
\input{privacy}

\section{Evaluation}
\input{evaluation}

\section{Conclusion}
\input{conclusion}

% \bibliographystyle{unsrtnat}
\bibliographystyle{IEEEtran}
\bibliography{marketplace}  %%% Uncomment this line and comment out the ``thebibliography'' section below to use the external .bib file (using bibtex) .

 \appendices
\input{appendix}

\end{document}

%% file: abstract.tex
Decentralized exchange markets leveraging blockchain have been proposed recently to provide open and equal access
to traders, improve transparency and reduce systemic risk of centralized exchanges. However, they compromise
on the privacy of traders with respect to their asset ownership, account balance, order details and their identity.
In this paper, we present \name{}, a fully decentralized privacy-preserving exchange marketplace with support for
matching trade orders, on-chain settlement and market price discovery.
\name{} provides confidentiality of order rates and account balances and unlinkability between traders and their trade orders, while retaining the desirable properties of a traditional marketplace like front-running resilience and market fairness.
We define formal security notions and present a security analysis of the marketplace.
%We also present \malicious{}, a version of the protocol which protects against malicious adversaries.
We perform a detailed
evaluation of our solution, demonstrate that it scales well and is suitable for a large class of 
goods and financial instruments traded in modern exchange markets.

%% file: intro.tex
\label{sec:intro}
Exchange markets have grown to be the cornerstone of trade and a measure of the
health of an economy. 
%Besides stock exchanges that permit trading in stocks and various other
%financial instruments, global commodity exchanges permit traders to buy and sell products ranging from
%agriculture, mining, oil, chemicals and energy. 
While modern marketplaces provide sophisticated services including regulatory oversight, historically they provided three critical functions. 
First, they matched buyers and sellers, previously unknown to each other, to enable trade. Second, they facilitated a
fair exchange between a buyer and seller by decoupling the exchange between them into two
contracts, %one each that the buyer and seller held with the central exchange (often through a clearing house). 
%This permitted the buyer and seller
such that they did not have to enter into a contract with each other, but instead each have a contract with the exchange which was much more
reputed and one with significantly lower counterparty risk of settlement. Third, the market permitted price discovery, enabling
traders to learn the current fair price determined by market economics, at which they could buy or sell goods.
In lieu of these services, the exchange would charge a transactional fee for each trade.

Over time exchange markets have grown significantly in size and volume and command high profit margins.
Complex cumbersome processes that traders must follow have created
an ecosystem of intermediaries who reduce transparency, increase costs and create higher barriers for entry.
%Fees may vary based on whether the trader is a member or a 
%participant in incentive programs, and whether the trader is a high frequency or long term trader.
%This presents a higher barrier of entry for smaller traders, who end up paying higher transaction fees due to
%lower negotiation power. 
The global commodities trading forum report 2015~\cite{Forum-Report} cites transparency
as one of the pressing concerns in the trading ecosystem. Further, centralized exchanges have also been the target of security attacks. 
A 2013 study~\cite{IOSCO-Study} reported that
53\% of exchanges surveyed had been hit by a cyber attack in 2012. A majority of these attacks were malware, denial of service
and information theft attacks, and no financial theft was reported. Several centralized cryptocurrency exchanges such as Mt. Gox~\cite{MtGox}
have been hacked in recent years involving theft of several millions of dollars in cryptocurrency.

In an effort to provide equal access to all traders, improve transparency, and reduce systemic risk associated with the
reliance on a trusted central entity for exchange operations, decentralized exchanges have been proposed in the recent past 
with regards to cryptocurrencies~\cite{Idex, airswap, BinanceDex, Bisq, 0xProtocol, BlockDX} and commodities~\cite{Affogato}. These decentralized exchanges leverage the cryptographic security
and immutability of blockchain to track asset ownership, orders placed by traders and execute settlement in a distributed and replicated
ledger. While they address many of the above issues with centralized
exchanges, they raise two important challenges which have caused them to not be practical in reality. 
% \cite{deX}

First, centralized exchanges 
implicitly provide confidentiality of information pertaining to traders' asset ownership, account balances, orders prices and
identity, with respect to other traders. The exchange itself acts as the sole and trusted custodian of this information, while only revealing 
information of recent trades in aggregate that is necessary for price discovery by the traders,
for example, the top-$K$ matched orders for some $K$.
Current decentralized exchanges
maintain this information in the clear on their ledgers, making it known to a larger set of entities that have a copy of the ledger.
Prior research~\cite{TransparentDishonesty-FC19, FlashBoys-Arxiv19, FrontRunning-FC19} has highlighted how public information regarding 
trades on blockchain have been leveraged in the past to launch front-running attacks. Further, financial privacy of asset ownership and trades is
a key requirement in use cases such as private equity trading markets~\cite{nasdaq-private-market}. Brokers who assist in private equity trades are expected to maintain privacy
of all financial transactions of their clients and only the centralized exchange is privy to the specific trades and their owners.
An exchange also needs to be fair~\cite{MaximalMatching-AAMAS13}, where less competitive orders cannot be matched, while more competitive orders are unmatched.
% Front-running attacks leverage details of ongoing orders and use it to economic advantage.
% For instance, a front-running buyer could use knowledge of competing buyers to out-bid them.
% A centralized operator could front-run by submitting its own order.

% While privacy-preserving payment networks such as~\cite{ZeroCash, ZKLedger, Solidus, Zether} have been proposed,
% % leveraging Zero Knowledge Proofs (ZKP) and blockchain,
% to our knowledge, our work is the first to provide fully
% decentralized privacy-preserving matching, settlement and price discovery services for an exchange market. 

Privacy-preserving payment networks on blockchain such as~\cite{ZeroCash, ZKLedger, Solidus, Zether} have been proposed, supporting confidentiality of transaction values along with user anonymity.
In payment networks, it is inherently assumed that the trading
parties know each other and have {\em agreed} for the transfer of certain assets or cryptocurrency. Zero knowledge proofs
are constructed as per this prior agreement to ensure the integrity of the transfer while providing confidentiality of the assets and unlinkability of the trades to the buyer and the seller. In exchange markets, no such prior agreement exists. 
In the buy and sell orders we consider, no particular recipient is specified and only a constraint on the price is indicated.
Moreover, to the best of our knowledge, while current state-of-the-art on decentralized privacy-preserving marketplaces provide front-running resilience which preserves order confidentiality until matching, they do not support order rate confidentiality throughout.
How the matching problem based on these privacy constraints on orders and the settlement problem based on private account balances and orders
are solved in a decentralized manner while providing unlinkability between the trades and the traders form the crux of this paper.
% We leverage a novel combination of blockchain, ZKP and secure multiparty computation, and provide a suite of solutions with varying trade-offs between privacy and performance.
Our fully decentralized solution without reliance
on a trusted custodian or operator provides matching fairness and resilience against front-running and denial of service attacks.
It also provides order rate confidentiality even after completion, revealing only minimal information necessary to support price discovery.
 
A second significant challenge of decentralized marketplaces today is scalability and performance. 
%Recent years have seen significant advances on this important research topic in all the building blocks we leverage, namely blockchain, 
%zero knowledge proofs and multiparty computation. 
% Our system requires only a one-time lightweight participation from traders, when submitting the order.
% While we do not directly improve upon the scalability of the underlying cryptographic techniques,
We demonstrate that our privacy-preserving
decentralized marketplace is practical and scales to more than a thousand orders per minute, 
which is more than the trading volume for several classes of commodities in some of the largest centralized exchanges today.
We envision that this performance will only increase with improvements in the underlying cryptographic building blocks.
Additionally, a trader in our system only needs a one-time lightweight participation per trade.
% to truly address the scaling needs of modern exchanges.

Our main contributions in this paper are as follows:
\begin{itemize}
\setlength\itemsep{0em}
\item We propose \name, a fully decentralized privacy-preserving exchange market with matching, settlement and price discovery services, 
providing confidentiality to the trade rates and the account balances and unlinkability between the trades and the traders.
\name{} also supports front-running resilience, market matching fairness and market availability, all while letting traders be arbitrarily malicious.
% \item We present a suite of alternative solutions that trade-off privacy and performance, and introduce a measure of privacy to quantify the information leakage.
\item We define formal security properties and present a security analysis of the marketplace, including a measure of privacy to quantify the information leakage (necessary to enable price discovery).
%  \item We present versions of the protocol that protect against semi-honest adversaries (\name{}) and malicious adversaries (\malicious{}).
\item We demonstrate through detailed experiments, that our proposed methods scale well and are suitable for a large class of commodities 
and financial instruments.
\end{itemize}

%The rest of this paper is organized as follows. 
We describe the system model and problem statement in Section~\ref{sec:problem}. We discuss 
relevant background and provide an overview of our solution in Section~\ref{sec:overview}.
We formally define the security properties of our system in Section~\ref{sec:security}.
Details of the \name{} marketplace protocol are 
presented in Section~\ref{sec:protocol} and in Section~\ref{sec:sec-analysis}, we analyse the security of our protocol, quantify its privacy and discuss availability attacks.
We discuss details of our evaluation in Section~\ref{sec:evaluation}
and conclude in Section~\ref{sec:conclusion}. 

%%% Local Variables:
%%% mode: latex
%%% TeX-master: "paper"
%%% End:

%% file: problem.tex
\label{sec:problem}
Our system model consists of a set of traders who wish to buy or sell an asset in a marketplace. The marketplace
performs the function of matching buyers and sellers, and facilitates exchange and settlement.
We assume that the traders place {\it limit orders}, which is a type of order to buy or sell the asset
at a specified price or better. That is, {\it buy orders} specify the maximum price the buyer is willing to pay and
{\it sell orders} specify the minimum price the seller is willing to sell at. For simplicity, we assume that each order
is for exactly a unit quantity of the asset (any order for multiple units of the asset can be considered as multiple
orders each of quantity one).
Such an assumption does not support features such as discounts for bulk orders and does not optimize for cost of delivery.
There could be an interesting future work for privacy-preserving matching with support for trade volume and privacy of asset quantity.
We use the terms order price and rate interchangeably. 

% Review: It is not completely clear to me what is the information about sellers that is actually recorded on the blockchain. It seems to me that the blockchain records the associations between assets and sellers. If so, then as the number of assets changes for a certain seller, one can learn information about the "business" of a certain seller.
Note that an asset can be any physical, digital or virtual asset.
and each order bid price may be specified in a 
standard currency that the marketplace operates in (this can be trivially extended to multiple types of assets that
are traded in the marketplace and to even exchange between two assets directly). 
We only concern ourselves with recording
the rightful ownership of the asset on a ledger managed by the marketplace after each trade and do not handle the physical exchange of assets.

All traders are registered and hold an account with the marketplace. We consider spot markets, where a trader must have sufficient balance in their account
for posting a buy order and we do not consider margin buying (a scenario where the buyer purchases an asset using borrowed funds from
a broker, with the purchased asset as collateral). A trader must hold the asset in order to post a sell order and we do not consider
short selling (a scenario where a trader expects the price of the asset to go down, so sells the asset first and then buys it back
within a specified time frame). 

In many blockchain implementations including Bitcoin~\cite{Bitcoin}, the UTXO (Unspent Transaction Output) model 
has been popular rather than an account model. In the UTXO model, each transaction consumes a set of UTXOs and creates new UTXOs,
possibly with a different set of owners.
This makes checking for double spending easier as no two transactions can consume the same UTXO. However, in real-world
scenarios and specifically for exchanges, the account
model is a lot more practical and beneficial as additional functions such as KYC (Know Your Customer), authorization, deposit interest rates and loan rates 
can be supported. Hence, we use the account model in this work.

The marketplace records all buy and sell orders received in an order book. Periodically, it triggers a matching algorithm
to match buyers to sellers such that their price constraints are satisfied, i.e., for each buyer and seller that are matched, the buyer's price
is no less than the seller's price. For matched orders, the marketplace also determines a settlement price to complete settlement,
wherein the buyer pays the seller the settlement price and the seller transfers ownership of the asset to the buyer. Orders 
not matched in one round are automatically carried forward to the next round of matching, up to a timeout after which they
are expelled from the order book. This is necessary to ensure that the order book does not get clogged with irrelevant orders that are 
unlikely to be matched.  

Traditionally, a centralized trusted organization with strict audit controls manages such a marketplace. Privacy is supported
for the traders, but not with respect to the marketplace owner, who has access to all information. In this paper,
we consider a decentralized marketplace of untrusted {\it facilitators} and endeavor to extend privacy to include the facilitators as well,
and not just from other traders.

%%% Local Variables:
%%% mode: latex
%%% TeX-master: "paper"
%%% End:

%% file: overview.tex
\label{sec:overview}
% We first provide background on certain key techniques and algorithms we leverage in Section~\ref{sec:background}
% and then describe various solution alternatives in Section~\ref{sec:alternatives}.

\subsection{Background}
\label{sec:background}
{\bf Pedersen Commitments:}
Commitment schemes are cryptographic techniques that allow one to commit to a value such that the commitment is both hiding and binding. 
The hiding property ensures confidentiality of the committed value, while the binding property ensures that the commitment cannot later be opened to another value.
Pedersen commitment~\cite{pedersen1991threshold} (henceforth referred to as PC or just commitment) 
is a widely used commitment scheme which is also additively homomorphic.

\noindent
{\bf Matching algorithm and Fairness:}
Many matching algorithms have been proposed in literature and are widely adopted by centralized
exchanges today~\cite{MatchingAlgorithms,MatchAlgorithmsCME}. Among them, the price-time algorithm is arguably the most popular.
Buy orders are sorted in decreasing order of price and sell orders in increasing order, so that the most
competitive orders by price are at the head of the lists. Orders are matched sequentially from the two lists until no more orders can be matched, that is 
the next buy order's price is less than the next sell order's price.
We use a notion of fairness introduced in~\cite{MaximalMatching-AAMAS13} in the context of double-sided auctions. A fair matching is fair to both buyers and sellers. 
A matching is fair to the buyers if there does not exist any unmatched buyer $B_1$, such that a buyer $B_2$ with order rate less than $B_1$ is matched. 
Similarly, the matching is fair to the sellers if there does not exist any unmatched seller $S_1$, such that seller $S_2$ with order rate more than $S_1$ is matched.
The price-time algorithm is fair as it will always matches the most competitive orders first. 

% \subsubsection{Multiparty Computation}
% Secure Multiparty Computation (MPC) is a field of cryptography which allows multiple untrusting parties to collectively compute a function of their private inputs 
% with the guarantee that no party learns any other party's inputs and either all parties or no party gets the final output.
%Suppose the function to be computed is $F$ and the inputs of each party is $x_1$ to $x_K$.
%Then the final output of all the parties is $F(x_1, ... , x_K)$ if the MPC is successful, else no party gets any output.
%MPC techniques can be broadly divided into two types based on the security properties: semi-honest (passive) or malicious (active) adverseries.
%Semi-honest adverseries will not deviate from the protocol, but will try to gain information from the system. Malicious adverseries can 
%arbitrarily deviate from the protocol.
%In both cases, MPC guarantees that the privacy properties hold. Malicious adversary protocols are generally less efficient due to the additional security 
%guarantees they must provide.
%All MPC protocols make certain assumptions regarding the parties, such as all parties being able to communicate with each other and no party
%fails during the protocol. 
% While the privacy and security guarantees of MPC ensure that no information is leaked during the computation, 
% the output of the MPC function itself can leak information. MPC has been demonstrated to be efficient and practical for usage in real-world applications~\cite{bogetoft2009secure, burkhart2010sepia,10.1007/978-3-642-32946-3_5}.

\subsection{Decentralized Exchanges on Blockchain}
The tamper-proof ledger of cryptographically signed transactions in a blockchain support the verifiability and accountability needed 
for decentralized exchanges.
Smart contracts, which allow trusted distributed computation of functions, guarantee correctness and allow marketplace functionalities 
to be carried out in a trusted and risk-free manner. 
%The tamper-proof ledger of cryptographically signed transactions ensures verifiability and accountability,
%enabling trusted and risk-free settlement.
There are a large number of decentralized exchanges today for trading cryptocurrencies. Exchanges such as 0x Protocol~\cite{0xProtocol}
and IDEX~\cite{Idex} adopt a centralized off-chain matching engine with on-chain settlement.
%\cite{RadarRelay}, \cite{Kosu}, \cite{DDEX}, Nash~\cite{Nash}, and Blockchain.io~\cite{blockchainIO}, and WavesDex~\cite{wavesdex}
Airswap~\cite{airswap} and Bisq~\cite{Bisq} expect traders to match each other 
(either off-chain or on-chain) by providing a peer-to-peer network for discovery and perform settlement on-chain and thus, do not offer automatic order matching. 
% Altcoin.io~\cite{altcoin},AtomicDEX~\cite{atomicdex} 
%Altcoin.io and AtomicDEX use atomic swaps enabling traders to trade across multiple blockchain 
%networks for trustless exchange of assets for settlement. 
Binance DEX~\cite{BinanceDex} and BlockDX~\cite{BlockDX} handle both matching and settlement on-chain.
None of these exchanges support privacy of account balances or orders and store all information in the clear on the blockchain.

Other related work in the space of privacy-preserving decentralized exchanges include ZEXE, TEX, FuturesMEX, Dark MPC and P2DEX~\cite{ZEXE, TEX, FuturesMEX-SP18, darkMPC, P2DEX}.
We provide a detailed comparison with our work in Table~\ref{table:comparison} with respect to the security properties we define in Section~\ref{sec:security}.

% \vspace{-0.1in}
\subsection{Design Alternatives}
\label{sec:alternatives}
We provide an overview of various alternatives for designing a marketplace that exercise
different trade-offs between performance and privacy. We categorize the alternatives across three dimensions 
of how information is shared, namely (i) level of decentralization of the marketplace, (ii) confidentiality of account 
balances, and (iii) confidentiality of order prices.

%Traditionally, marketplaces have relied on a trusted third party to centrally manage the operations of the market.
In recent years, decentralization of the marketplace has been proposed leveraging blockchain to reduce reliance on the trusted party,
and improve security, transparency and operational costs by mitigating counterparty risk. 
%Blockchain,
%with its ability to store information in a secure, immutable, and decentralized manner, has been utilized to 
%address the problem of exchanging assets between distrusting traders without a trusted intermediary~\cite{Idex,Bancor,OasisDEX,Affogato,deX}.
However, an adverse side effect with decentralization is the additional loss of privacy. Traders' information,
including their account balances and order details are now public to the decentralized set of entities operating the
marketplace and anyone having access to the blockchain ledger. 
%In a bid to support complete privacy of account balances, 
Prior work in supporting confidential payments on blockchain~\cite{ZKLedger} has supported privacy of account balances leveraging ZKP.
%have leveraged commitment schemes such as Pedersen commitments~\cite{pedersen1991threshold}. 
%While these schemes are computationally much slower than sharing values in the clear, they have increasingly become
%practical over the years.

As motivated earlier, it is not sufficient for only the account balances to be private. To prevent front-running and other attacks based on information theft, the orders placed
by traders also needs to be protected. When account balances are public
information, it does not help to keep the orders private as anyone who views the account balances before and after a transaction
can discern the specific order price. So, we only consider protecting order information when the account balances are also private.

In order to match buyers and sellers, the marketplace needs to know if the price quoted by a seller is lesser than the
maximum price a buyer is willing to pay. In other words, the marketplace needs to be able to perform comparison between
buyer and seller prices to match orders. A central contribution of this paper is to support such a matching algorithm with encrypted prices and guarantee reliable 
and automated settlement in a decentralized manner.
% We consider two main design alternatives to support this -- one that is
% more performant but only supports partial privacy, and another that supports complete privacy for the traders under certain trust assumptions.

One approach that supports partial privacy is inspired by the classic bucket sort algorithm~\cite{cormen2001section}. 
The main idea of the bucketization based protocol is that the rate range is split into non-overlapping buckets, trading privacy for performance.
% TODO: mention the 2 phases or stick to one phase
% Trader submit orders with private rates as commitments along with the public bucket values that their rates fall in.
Traders submit orders in two phases, wherein they submit their private rates as commitments in the first phase. The marketplace chooses random bucket 
start and end values after which traders disclose the corresponding public bucket values that their committed rate belongs to, in the second phase,
along with a ZKP of range. 
This provides a coarse sorting of orders and the marketplace subsequently matches buyers in higher price buckets with sellers in lower price buckets.
Matched orders are settled at the respective private rates of traders with additional zero knowledge proofs to guarantee integrity.
We describe the order submission, matching and settlement of orders in this protocol in greater detail in Appendix~\ref{sec:appendix-bucketization}
and use it as a baseline for comparison in our experiments in Section~\ref{sec:evaluation}.

We propose \name{}, a privacy-preserving marketplace protocol. Secure multi-party computation (MPC) is one of the building blocks of our protocol.
As MPC may not scale well to a large number of traders, and traders cannot be expected to always be online and reliably carry out 
computations needed by MPC algorithms, we introduce a set of {\em brokers}, fixed in number, to which the traders secret share the prices and enable them perform the necessary computations on private data. %Traders secret share their prices to the brokers, who collectively add up to their order price. 
The MPC properties ensure that the individual brokers have no information regarding an order's price under reasonable collusion assumptions. 
%We show how matching and settlement of orders can be done privately in this method in Section~\ref{sec:mpc-protocol}. 
Note here that a broker is a
role performed by an entity. It is completely feasible that the facilitators operating the peers of the blockchain for the marketplace
are themselves brokers, but for ease of exposition we describe them as separate entities from the blockchain peers. We refer to the 
brokers and the blockchain peers collectively as the marketplace.

\begin{figure}
    \centering
      \includegraphics[width=0.6\linewidth]{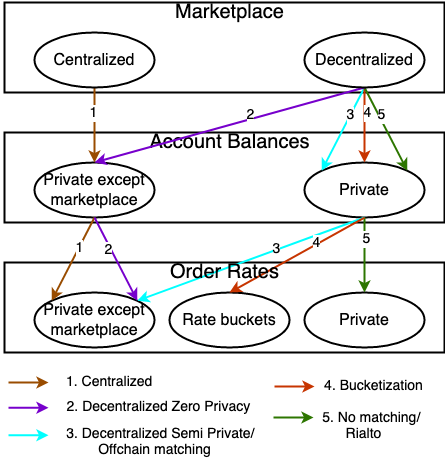}
    \caption{Design alternatives for a marketplace}
    \vspace{-0.1in}
    \label{fig:overview}
\end{figure}

Figure~\ref{fig:overview} pictorially represents the various alternatives mentioned above. The \textsc{Centralized} marketplace 
has a trusted third party operating the marketplace. Traders share their accounts and orders openly with the marketplace operator, who
is trusted to maintain it privately, denoted by `Private except marketplace'. The \textsc{Decentralized Zero Privacy} mechanism replaces the trusted
third party with a set of facilitators operating a blockchain, which is used to store account balances and orders. Here again, account
balances and orders are private except to marketplace facilitators. The next alternative is \textsc{Decentralized Semi-Private}, where
account balances are maintained as commitments on the blockchain and not revealed to even the marketplace facilitators, but order details
are shared with the marketplace in the clear. For both the above
decentralized designs, it is possible to perform the matching component outside the blockchain by a trusted third party and only record 
the results on blockchain. This helps achieve greater performance at the cost of trusting the entity performing the off-chain matching. 
We consider this variant \textsc{Offchain Matching} as many decentralized exchanges today such as~\cite{Idex,0xProtocol} adopt this.
Both the \textsc{Bucketization} scheme and \name{} support orders to
also be private from the facilitators.
% In order to characterize the overheads of sorting using MPC and the matching on-chain in \name{}
% , we consider a variant where there is no matching and only settlement, \textsc{No Matching}.
% Here, we assume that traders 
% agree with one another for an exchange and leverage the marketplace only for reliable settlement.

%%% Local Variables:
%%% mode: latex
%%% TeX-master: "paper"
%%% End:

%% file: security.tex
\label{sec:security}
%We define confidentiality and anonymity properties of a trader even when all other traders in the marketplace collude.
We now define the security properties of our system. Let $\secp$ be the security parameter. %The definitions are with the assumption that traders in the marketplace do not collude among each other, but they can be modified such that the security of the system degrades gracefully when a subset of traders collude against the others.
We have the marketplace operate in rounds with trades received up to a certain (possibly periodic) timer considered for matching in each round. Let $N$ be the total number of traders in the marketplace. Let $N_i$ be the total number of traders participating in round $i$ and let $T_i$ be the set of traders $\{t_{ij}\}_{j \in [N_i]}$ participating in round $i$. %Let $\{\bar{t}_{ij}\}_{j \in [N_i]}$ be temporary ids assigned to $\{t_{ij}\}_{j \in [N_i]}$.
The view of a trader in the marketplace at the end of round $i$ consists of the entire ledger till round $i$, its account balance at the beginning of each round $i' \in [i]$ and its order rate for each round $i' \in [i]$. We define the security properties from the perspective of a single malicious trader. These properties can be extended to capture a collusion among multiple malicious traders.
%\begin{itemize}
%\item its account balance at the beginning of round $i'$
%\item its order rate for round $i'$ (buy rate for a buyer or sell rate for a seller)
%\item the entire ledger till round $i'$
%%\item for each trader in $T_{i'}$, the round they last participated in
%%\item ordering (inequalities) between the order rates of the participating traders $T_{i'}$
%%\item top-$K$ order rates in round $i'$
%%\item set of commitments $\{c^b_j\}_{j \in N}$ of account balances of all traders at the end of rounds $i'$  
%%\item set of commitments $\{c^o_j\}_{j \in N_i}$ of order rates for the $N_{i'}$ traders participating in round $i'$
%\end{itemize}
%$\buyers_i$ be the set of buyers at 

\noindent {\bf Property 1:} {\it Confidentiality of Orders:} %No information about the order rate of a trader is revealed before the matching protocol begins. 
No information about the order rates except the top-$K$ settlement rates is revealed at the end of a trading round.

\begin{definition}[Order rate confidentiality]
There exists a probabilistic polynomial time (PPT) simulator $\Sim$ such that for every round $i$ and for every adversarial trader $t_{ij}$ in $T_i$, $\Sim$ simulates the view of $t_{ij}$ when provided with (i) the top-$K$ order rates for rounds $\leq i$, (ii) the account balance of  $t_{ij}$ at the beginning of round $i$ and (iii) the order rate of $t_{ij}$ for rounds $\leq i$. The probability for any PPT algorithm $\Adv$ to distinguish between the real view and the simulated view is $1/2+\negl(\secp)$. 
\end{definition}

We use the notion of a \textit{relaxed order rate confidentiality} for our system where the simulator is provided the \textit{order book leakage}, which includes all the protocol specific leakages in addition to the top-$K$ settlement rates.

%\begin{definition}[Relaxed order book confidentiality]
%There exists a PPT simulator $\Sim$ such that for every round $i$ and for every trader in $T_i$, $\Sim$ simulates the view of the trader when provided with (i) the ordering between the order rates in round $i$, (ii) the top-$K$ order rates for round $i$, (iii) the account balance of $T_i$ at the beginning of round $i$ and (iv) the order rate of $T_i$ for round $i$.
%\end{definition}

\noindent {\bf Property 2:} {\it Unlinkability of trades with traders:} 
An order cannot be linked to a trader who placed that order. Also, orders from different rounds cannot be linked back to the same (potentially unknown) trader who placed them. We formalize this property as follows.
%In \name{}, the only information revealed about the account associated with an order is the round that the same account participates in next. If all the traders participate in each round, \name{} reveals no information about the orders associated with a trader.
\begin{definition}[Trader unlinkability]
For any pair of rounds $i$ and $i'$, the probability of linking an order in round $i$ placed by a trader in $T_i$ to an order placed by the same trader in round $i'$ is at most $1/min(N_i,N_{i'})$.
%For any trader and for any pair of rounds $i$ and $i'$ that this trader participates, the probability of associating a trade in round $i$ and a trade in round $i'$ to this trader is at most $1/min(N_i,N_{i'})$.
\end{definition}
%\vdnote{1) Is this a definition?, 2) Is this the strongest that we can claim?}

\noindent {\bf Property 3:} {\it Confidentiality of Traders' Accounts:} No information is revealed about the account balance of a trader except the information inferred from the union of order book leakage across rounds.
\begin{definition}[Account balance confidentiality]
We have an indistinguishability based security game between two PPT algorithms: a challenger $\Chal$ and an adversarial version $\Adv_t$ of trader $t$. 
\begin{itemize}
\setlength\itemsep{0em}
\item $\Adv_t$ picks a trader $t^*$, and two values $b_1$ and $b_2$. 
%\item $\Adv$ picks a trader $t^*$, initial account balances for all $N$ traders except $t^*$, order rates and two values $b_1$ and $b_2$ which incur the same order book leakage. 
\item $\Chal$ randomly chooses one among $\{b_1, b_2\}$ as the starting account balance for $t^*$ in the first round it participates in. The marketplace proceeds with $A_t$ orchestrating $t$ and $C$ orchestrating $t^*$.
\item $C$ aborts the game if the allowed order book leakage distinguishes between $b_1$ and $b_2$. 
\item $A_t$ eventually outputs its guess $b_1$ or $b_2$.
\end{itemize}
The marketplace is said to have account balance confidentiality if the probability for $\Adv_t$ to correctly guess $\Chal$'s choice when $\Chal$ does not abort is $1/2 + \negl(\secp)$.
\end{definition}
%The definition intentionally does not quantify the \textit{rounds in which the trader's participation is revealed} since this varies with the system.

\noindent {\bf Property 4:} {\it Front-Running Resilience:} A malicious trader cannot learn information of an ongoing unmatched order to submit an order of their own and gain economic advantage.% No information about the order rate of a trader is revealed before the matching protocol begins. This is formally defined in the same way as order rate confidentiality but with the simulator provided the order book leakage only till round $i-1$.\footnote{We provide different definitions for order rate confidentiality and front-running resilience since the view of the adversary is different in these two settings.}

Front-running resilience is ensured by the confidentiality of an order from other traders before the matching phase begins.
Order rate confidentiality deals with hiding the order rates, even after matching and settlement are completed.
While front-running resilience is supported by most marketplaces, order rate confidentiality is a more general privacy property to ensure only minimal necessary information about order rates is revealed, for price discovery.

\noindent {\bf Property 5:} {\it Market Matching Fairness:} A less competitive buy or sell order (a buy order with a higher bid price or a sell order with a lower ask price) cannot be matched, when a more competitive order remains unmatched.

\noindent {\bf Property 6:} {\it Market Integrity:} Once a buy and sell order are matched, the marketplace automates settlement. A trader attempting to deviate from this can be identified and penalized. Market integrity establishes the following:
\begin{itemize}
\setlength\itemsep{0em}
\item Buy rate is greater than or equal to the sell rate in each settlement. And there is no settlement which induces a buyer to pay more than its buy rate and a seller to be paid less than its sell rate.
\item Each settlement ensures that the amount deducted from the buyer exactly equals the sum of amounts credited to the seller and to the marketplace as fees.
\end{itemize}

%\noindent {\bf Property 7:} {\it Price Discovery for Traders:} The marketplace reveals the top-$K$ matched orders so as to permit traders to discover current market price for an asset.
%
\noindent {\bf Property 7:} {\it Market Availability and DoS Resilience:} The marketplace is designed to be resilient to Denial-of-Service and other attacks on its availability.

Our privacy-preserving decentralized marketplace enables automatic settlement and price discovery for traders in addition to the above security properties. 
% Price Discovery permits traders to discover current market price for an asset by revealing the top-$K$ matched orders.

%\subsubsection*{Market integrity}
%Money integrity and matching correctness
%
%\subsubsection*{Money stability}
%No loss or gain of assets in marketplace between the beginning a round and its end.
%\begin{itemize}
%\item There exists a balance $B$ at the beginning of the marketplace such that $\sum_i \buyerBal_i +  \sum_i \sellerBal_i + \marketBal = B$ at the end of each round. \vdnote{Check}
%\end{itemize} 
%
%\subsubsection*{Matching correctness}
%In every match, $\buyRate \geq \sellRate$.
%
%\subsubsection*{Trader solvency}

%%% Local Variables:
%%% mode: latex
%%% TeX-master: "paper"
%%% End:

%% file: protocol.tex
\label{sec:protocol}

\begin{figure}[ht]
    \centering
      \includegraphics[width=1.0\linewidth]{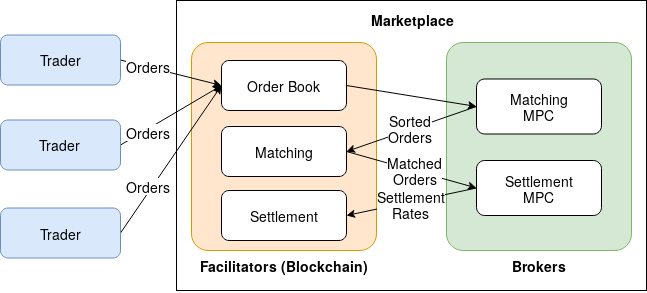}
    \caption{Architecture diagram of the different components of the system for the \name{} protocol.}
    \label{fig:arch}
\end{figure}

An exchange marketplace has three main steps: order submission, matching and settlement.
\name{} uses a blockchain ledger that maintains commitments of the account balances of all participating traders.
At a high level, traders submit their orders by providing a commitment of their order rates to the ledger, and secret share their order rates to the brokers who perform a fair and maximal matching using MPC and a smart contract execution. We decide on an efficiency vs confidentiality tradeoff to move a part of the matching step outside MPC in a smart contract.
\name{} supports price discovery which reveals the top-K order rates after settlement. So, we provide an analysis of this leakage in conjunction with that of the tradeoff. Settlement is done in a smart contract using the homomorphic property of the commitment scheme while retaining the confidentiality of order rates.
Further, we make an efficient use of Waksman~\cite{waksman} network after settlement to perform an oblivious shuffle and achieve unlinkability of traders. During these steps, to protect against malicious brokers from modifying the input shares from the traders, we deploy an efficient distributed version of the Schnorr protocol~\cite{schnorr} linking the shares to the corresponding commitments on the ledger.
 The components of \name{} are shown in Figure~\ref{fig:arch}.

\subsection{Trust model}
\label{sec:trust-model}
We make the following trust assumptions on the entities in our system:

\noindent {\it Traders:} We assume all traders to be untrusted and financially incentivised to lie about asset ownership, account balances and attempt to cheat the system.

\noindent {\it Blockchain facilitators:} While individual facilitator nodes can be malicious, we assume
a sufficient majority of them to be honest to preserve the integrity of the blockchain (e.g., two-thirds honest majority when using byzantine-tolerant consensus). The adversarial blockchain nodes may also collude with the traders.

\noindent {\it Broker nodes:}
A broker node may collude with the traders, blockchain facilitators or other broker nodes.
Depending on the requirements of the marketplace, adversarial broker nodes could be in minority or in majority.
We present \name{}, a version of the protocol where the adversarial broker nodes are semi-honest (try to gain information without deviating from the protocol) and \malicious{} where they are malicious (can arbitrarily deviate from the protocol).
Note that we treat the colluding traders and blockchain facilitators to be malicious in both these versions.

\subsection{Protocol Description}
\label{sec:mpc-protocol}
\begin{figure}[ht]
    \centering
      \includegraphics[width=1.0\linewidth]{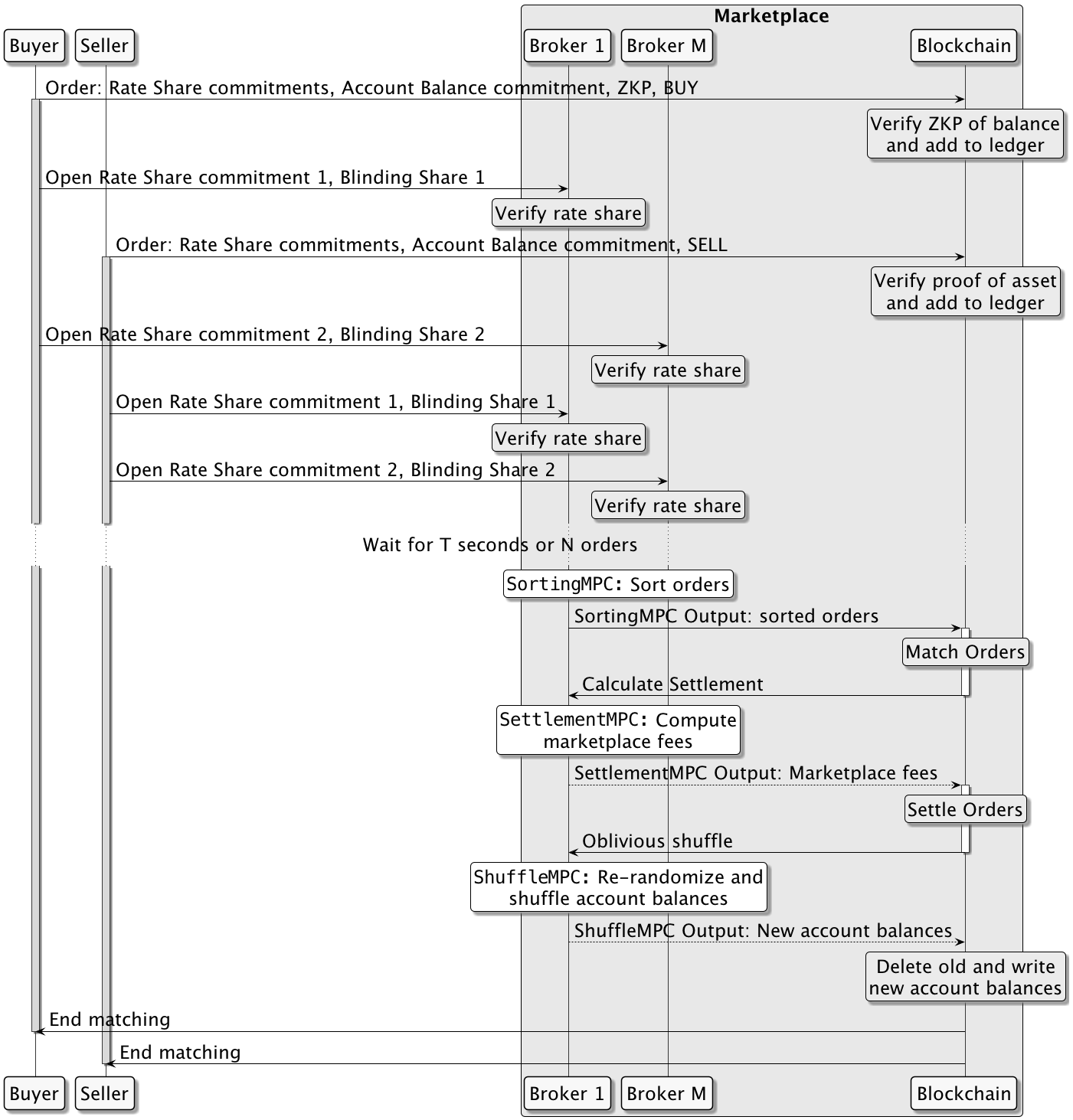}
    \caption{Sequence of calls for the \name{} protocol.}
    \label{fig:MPCsharing}
\end{figure}

Figure~\ref{fig:MPCsharing} provides a sequence of steps performed by the different entities involved in \name{}. Detailed algorithms are provided in Appendix \ref{sec:appendix-algorithms}.

\textbf{Order submission:}
Each trader secret shares its private rate value to get M additive shares, where M is the number of brokers in the system. 
The trader submits an order on the blockchain ledger containing commitments to each share.
A buyer also submits a commitment of its account balance (as stored on the ledger) and a ZKP of range~\cite{bunz2018bulletproofs} that its order rate is less than or equal to its account balance. This ZKP is verified by a smart contract.
Account balances are stored directly as commitments on the ledger, only the owner (which knows the value and blinding factor of the commitment) can submit a valid range proof.
The trader also submits a share of the rate to each broker by opening the commitment of this share submitted on the blockchain.
This secret sharing can be independent and outside of the MPC protocol used later by the brokers.
Additionally, the trader submits a share of a random blinding factor to the brokers, which will be used to re-randomize the commitments of its account balance, to be described in Section~\ref{sec:shuffle}.
% TODO: Says minority of brokers are malicious
In \malicious{}, secret sharing scheme is decided based on trust expected from the broker nodes. For instance, 
we use a $(k,n)$-threshold secret sharing scheme~\cite{secretsharing}, where $k = \lfloor n/2 \rfloor + 1$, to allow the protocol to proceed when a minority of the brokers are malicious and can abort.
For a buy order, the blockchain locks funds equalling the buyer's rate in an escrow by using the commitment of the rate, 
computed by ``adding'' the commitments of its shares. The commitment of the buyer's account balance is also ``subtracted'' with the locked value.
The marketplace allows only one trade order per account in a round, to prevent double spending.\footnote{Multiple trade orders per account can be permitted, by the trader submitting the orders along with a ZKP that (i) sum of their orders is less than or equal to the account balance and (ii) each of their orders is greater than or equal to 0. }

\textbf{Sorting using MPC:} The brokers wait for the earlier of T seconds or N orders and then perform MPC to determine the sorted list of orders.
The \SortingMPC\ algorithm takes N private inputs from each broker, where each input is a share of a trader's rate.
The \SortingMPC\ algorithm {\it securely} recomputes the shares and sorts the buy and sell orders together to output a single totally ordered list in an increasing order of prices.
% The privacy properties of MPC ensure that no broker learns any other broker's inputs.
The information leaked as a result of this step is the sorted list of all buy and sell orders, revealing relative positions.
The brokers finally submit the sorted list of orders to the blockchain smart contract, triggering the matching algorithm. 

A malicious broker, colluding with a trader, can modify traders' rate share inputs, gaining it an advantage during matching.
To prevent this, the brokers run the input share validation algorithm in Section \ref{sec:validation} to validate the rate share inputs used in \SortingMPC\ with the commitment of the rate shares which they obtain from the ledger.
%Brokers first perform the offline preparation phase of the validation algorithm $\mathcal{V}$.
%and then execute the online phase for all traders' rate shares. Then they run the offline verification phase and continue with the \SortingMPC\ algorithm with the honest subset of brokers.
On completing the validation algorithm, the brokers continue with the \SortingMPC\ algorithm using the validated shares.
  
\textbf{Matching:} After receiving the sorted list of orders from the brokers, a smart contract on the blockchain performs matching.
The algorithm considers the buy orders and sell orders in ascending order. A buy order is matched with the first sell order whose rate is lower than the buyer's rate.
This produces a maximal matching, the proof of which is presented in Appendix~\ref{appendix:matching-proof}.
Trade orders that have not been matched in a round are carried forward to the next round. To prevent flooding of the 
marketplace with unmatched orders, we remove orders that have not been matched for more than a fixed number of rounds.

\textbf{Settlement:}
For settlement, we use a scheme of type \textit{Marketplace earns the rate difference}.
This method of settlement is typically used as auctioneer's revenue in some
double auctions~\cite{10.1007/3-540-45749-6_34,wang2016truthful}.
In this settlement scheme, each trader settles at their proposed rate and the difference between the buyer and seller rates is paid as fees to the marketplace.
We settle between the matched buyers and sellers using the private proposed order rates as follows:
% The marketplace needs to compute the difference between the private rate values of all matched buyers and sellers.
% This is done using the brokers for MPC as follows:
\begin{enumerate}
\setlength\itemsep{0em}
\item The marketplace invokes the brokers for computing the marketplace fees.
\item The brokers run the \SettlementMPC\ algorithm with the rate shares of buyers and sellers as private inputs, and output a sum of all differences between buyer and seller rates. 
  Note that the difference in rates between a specific matched buyer and seller is not revealed by this computation, but only the aggregate across all matched pairs of orders in a round is computed. If required, this can even released as an aggregate across multiple rounds.
  % This can be made a private noisy release. And this release is also part of the order book leakage.
\item The brokers submit the aggregate value to the marketplace, triggering the settlement algorithm on the blockchain.
\item The smart contract settles matches by adding the commitment of seller's proposed rate to the 
  seller's account balance, enabled by the homomorphism of the commitment scheme.
  It also removes the funds locked in the escrow on behalf of the buyer. 
  The fees are added to the marketplace account.
\end{enumerate}

The traders to be settled are a subset of the traders who submitted orders. For \malicious{}, the input share validation performed before \SortingMPC\ will ensure validation of inputs for \SettlementMPC\ when continued with the validated shares.
% Similar to the \SortingMPC\, malicious brokers could modify the rate shares of one or more traders to either increase or decrease the marketplace fees, affecting market integrity.
% In \malicious{}, to protect against malicious adversaries, brokers run the offline preparation phase of the validation algorithm $\mathcal{V}$, then perform the online phase during MPC followed by the offline verification. 
% Finally, they continue with the \SettlementMPC\ algorithm with the honest subset of brokers.

After settlement, the difference between the old and new account balances of a trader equals the commitment of its order rate.
This permits all parties to associate an account to its corresponding trade order, leading to
linkability of accounts across rounds which affects account balance confidentiality and trader unlinkability.
In section \ref{sec:shuffle}, we propose a re-randomization and shuffling algorithm which occurs at the end of every round to un-link traders from accounts.

\subsection{Re-randomization and shuffling}
\label{sec:shuffle}

We propose a protocol using brokers to perform re-randomization and oblivious shuffle to achieve trader unlinkability.
Re-randomization with an oblivious shuffle of the commitments prevents all parties, including brokers and blockchain facilitators, from associating accounts to trade orders by matching the difference of balance to order rates.
%Oblivious shuffle of the account balances prevents both brokers and blockchain facilitators from associating accounts before and after re-randomization by shuffling the set of all input account balances.

This is done as follows. Traders submit a secret share of a random blinding factor, which will be added to the randomness in the commitment of their account balance, for re-randomization. This blinding factor is submitted during order submission, so the protocol does not require participation from the traders during this phase.
For oblivious shuffle, we use the Waksman permutation network~\cite{waksman}. When adhering to the setup in the rest of the protocol, we find this admits a better separation between the steps required to be performed inside the MPC and the steps brokers can perform outside and offline.
Waksman network is a permutation circuit with N inputs and N outputs along with control bits which decide whether specific pairs of elements should be swapped.
Other permutation networks such as Benes network have been shown to produce biased permutations~\cite{abe2001remarks} and will require multiple rounds to generate uniformly random permutations.
Waksman network works by choosing a permutation uniformly and computing the corresponding control bits, which can be done offline by each broker prior to the MPC.

% Given a permutation, there exists a recursive algorithm to obtain the control bits for the permutation network.
% Each broker chooses a random permutation locally and computes the control bits offline.
For each trader, each broker calculates the commitment of 0 using the blinding factor share offline.
Brokers perform the following \ShuffleMPC\ algorithm
with the above commitments and control bits of their permutation as private inputs and trader account balance commitments as public inputs.

\begin{enumerate}
  \setlength\itemsep{0em}
  \item For each trader, brokers reconstruct the commitment of 0 with the blinding factor from submitted commitment shares and ``add'' it to the commitment of the trader's account balance.
  % \item For each trader, brokers reconstruct the blinding factor from submitted shares, compute commitment of 0 using reconstructed blinding factor and add it to the commitment of its account balance.
% \item For each trader, brokers compute commitment of 0 using submitted blinding factors and add to corresponding account balances.
\item Output of previous step is permuted by composing each broker's permutation network.
\end{enumerate}

Finally, brokers output the re-randomized and shuffled account balance commitments to the marketplace.
Marketplace removes the old account balance commitments stored in ledger and replaces with the new set. This does not affect marketplace integrity since the account balance values themselves do not change during the protocol. Also, a trader can compute the new commitment locally since the re-randomization is done using the randomness it submitted.

For N inputs and M parties, a single Waksman circuit is of depth $O(\log N)$ with $O(N \log N)$ swappers. On composing $M$ networks, the total complexity of the shuffle is $O(MN \log N)$.

As before, malicious brokers can modify the blinding factor shares input to the protocol. This could prevent a trader from accessing its accounts in the future.
For securing against malicious brokers, the integrity of the blinding factor shares used is validated using the input share validation algorithm before proceeding with the \ShuffleMPC\ algorithm . %in the same way that the order rate shares were validated in the \SortingMPC\ algorithm.
%Brokers run the offline phase of validation algorithm $\mathcal{V}$, perform the online phase of $\mathcal{V}$ as MPC and then perform the offline verification.
%Finally, the honest brokers continue with the re-randomization and oblivious shuffle.
As described in Section~\ref{sec:shuffle}, brokers also provide control bits for the Waksman network to permute the traders' account balances.
A malicious broker may not input control bits for a uniformly random permutation network.
However, honest brokers are guaranteed to perform a uniformly random permutation which when composed with other brokers' permutations will be uniformly random.

\subsection{Input Share Validation Algorithm}
\label{sec:validation}
Malicious brokers may not only deviate from the MPC protocol, but could also modify the inputs to the MPC algorithm.
In \name{}, the private input shares to the MPC algorithms are grounded with their Pedersen commitments. The input share validation algorithm validates the input shares to the MPC by generating a ZKPoK of opening the commitments in a distributed manner.
These proofs can be verified outside the MPC protocol avoiding exponentiation inside the MPC algorithm to validate the input shares.

Consider $N$ secret values $v^t$ $\forall t \in [N]$, whose commitments using blinding factor $r^t$ are $c^t = g^{v^t}h^{r^t}$ and available to everyone.
Each value $v^t$ is secret shared along with its blinding factor $r^t$ among $M$ brokers, such that $v^t_i$ and $r^t_i$ are the shares for broker $i$ and commitment $c^t_i = g^{v^t_i}h^{r^t_i}$ is public.
Brokers would like to run an MPC algorithm $\mathcal{A}$ which takes as input the shares of secret values $v^t$ from each trader.

We describe an input share validation algorithm $\mathcal{V}$ which will validate each broker's input according to its public commitment before executing $\mathcal{A}$.
The proposed algorithm will involve an offline preparation phase, a validation phase involving an MPC and an offline verification phase.

\textit{Preparation phase:}
Each broker $i$ generates random $y^t_i, s^t_i \leftarrow \mathbb{Z}_q$ and calculates $d^t_i = g^{y^t_i}h^{s^t_i}$ locally for each trader $t$.

\textit{Validation phase:}
The following steps are performed as MPC, taking $y^t_i$, $s^t_i$ as private inputs and $d^t_i$ as public input apart from $\mathcal{A}$'s private inputs of $v^t_i$ and $r^t_i$.
%\begin{enumerate}[leftmargin=*]
%  \setlength\itemsep{0em}
%\item 
For each trader $t$,
\begin{enumerate}%[leftmargin=*]
    \item For each share, calculate $e^t_i = Hash(g, h, y^t_i, s^t_i)$
    \item For each share, calculate $a^t_i = y^t_i + e^t_i\cdot v^t_i$, $b^t_i = s^t_i + e^t_i \cdot r^t_i$
    \item Output $\{a^t_i, b^t_i, d^t_i, e^t_i\}_{i \in [M]}$
\end{enumerate}

Each broker then performs the following {\it offline verification phase}:
    \begin{enumerate}%[leftmargin=*]
    \setlength\itemsep{0em}
  \item Consider an $M$ bit vector $H = 1^M$.
  \item For each trader $t$ and for each broker $i$, $$H[i] = H[i] \wedge \left( g^{a^t_i}h^{b^t_i} = d^t_i(c^t_i)^{e^t_i} \right)$$
\end{enumerate}

On completing the verification, brokers continue with execution of algorithm $\mathcal{A}$ possibly with the brokers $i$ such that $H[i] = 1$. Blockchain can be used to get a consensus on this set.

This distributed proof generation clearly doesn't affect the soundness of the generated proof since the view of the ``verifier'' subsumes the view of a verifier in the non-distributed version.
Confidentiality of the distributed proof generation follows from the properties of the MPC protocol and the zero-knowledge property of the proof generation algorithm, along with the hiding property of the commitment scheme assumed for the non-distributed version.

\subsection{Matching Fairness}
\label{sec:fairness}
Our proposed matching algorithm is fair to sellers, since sellers are considered in ascending order of rates. However, it is not fair to buyers, since buyers 
are also considered in ascending order of rates. This makes it possible that buyers with the most competitive price quotes are left unmatched.

We propose a swapping of matched buyers with any unmatched buyers to introduce fairness in our matching without affecting its maximality.
% , in Algorithm~\ref{algo:fair-swapping}.
This is similar to the swapping procedure in~\cite{MaximalMatching-AAMAS13}. The algorithm swaps 
unmatched buyers with matched buyers lesser than it, by giving precedence to buyers with higher rates. Since after each swap the matches
continue to remain feasible and there are at most $n$ swaps, where $n$ is the cardinality of the maximal matching, the final matching
is both maximal and fair.

\subsection{Price Discovery}
\label{sec:price-discovery}
Typically marketplaces reveal certain statistics regarding the settled orders such as top-$K$ settlement rates at the end of each round, 
which helps in price discovery for buyers and sellers for subsequent rounds.
$K$ is usually a small percentage of $N$, the total number of orders in the round.
Price discovery is important, since it allows future traders to determine their limiting rates.
We explain how \name{} computes the top-$K$ settlement rates at the end of each round.
Our modified fair and maximum matching algorithm after swapping has the property that among matched buyers and sellers
$(B_1,S_1)$ and $(B_2,S_2)$, if $B_1$ has a higher order rate than $B_2$ then $S_1$ is no less than $S_2$.
%the highest price buyers are matched with the higher price sellers among all matched sellers.
This implies that the top-$K$ settlement rates are for the matches of the top-$K$ buyers, where the settlement rate is considered as the 
buyer's proposed rate for the scheme where marketplace earns the difference.
% , or computed as the mean of the traders' rates).
Where order rates are shared publicly with the marketplace, the marketplace facilitators can easily compute the settlement rates of 
the $K$ highest buyers and their matches and reveal it.
In \name{}, we do it by letting the MPC reveal the buyer's rate value (by reconstruction of shares) for each of the top-$K$ buy orders (after sorting) which have matched sell orders.

%%% Local Variables:
%%% mode: latex
%%% TeX-master: "paper"
%%% End:

%% file: privacy.tex
\label{sec:sec-analysis}
In this section, we analyse the security of our protocol. 
%the \textsc{Bucketization} and \textsc{MPC with Rate Shares} protocols with respect to the identity of traders, account balances 
%and the order book. 
In Section~\ref{sec:confidentiality}, we discuss how confidentiality of traders accounts and order rate is maintained (Properties 1 to 4 in Section~\ref{sec:security}).
We also discuss the integrity of \name{} marketplace (Property 6 in Section~\ref{sec:security}) in Section~\ref{sec:integrity}.
In Section~\ref{sec:quantify}, we quantify order rate privacy based on the information leaked by the system to measure the privacy of the protocols against marketplace participants.
In Section~\ref{sec:discussion}, we discuss availability and DoS resilience of \name{}.
In Section~\ref{sec:comparison}, we compare our work with related work on privacy preserving decentralized exchanges, along the axis of our security properties.

% ToDo: what is the *privacy notion* that you are trying to achieve with your proposed protocols. 
% ToDo:  including the threat model, a (cryptographic) definition of privacy, 
% ToDo:  adding a formal proof of that your protocol achieves the privacy notion that you have previously defined. 

% ToDo: Make this section as the main security analysis
% A. Threat analysis
% B. Order Book Guarantees
% 1. Front-Running Resilience
% 2. Trade Confidentiality

% \begin{subsubsection}{MPC with Rate Shares}
\subsection{Confidentiality and Unlinkability}
\label{sec:confidentiality}
We argue the confidentiality and unlinkability properties of our system at a high level in this section. As mentioned in Section \ref{sec:security}, the properties are defined with respect to a single malicious trader which can collude with the adversarial blockchain and broker nodes and these can be extended to capture a collusion among multiple malicious traders. 

The high-level arguments for the claims presented in this section also assume semi-honest broker nodes. We already expect the semi-honest broker nodes to collude with the malicious blockchain nodes and traders. The colluding traders can reveal their order rates and account balances to the broker nodes, while the blockchain nodes do not input secret information in any of the rounds. Hence, the extension of the arguments to malicious broker nodes involves proving that the broker nodes cannot use the colluding traders' order rates and account balances to deviate from the protocol and gain more information about honest traders than what's gained by the semi-honest broker nodes colluding with the malicious traders and blockchain nodes. An observation that will help here is that the inputs to the broker nodes are secret shares of all the private inputs from the traders and their commitments, and hence the collusion with a subset of traders reduces the MPC to that with inputs as the shares from the honest traders.

\noindent {\bf Order book leakage:} The consolidated order book leakage in \name{} for round $i$ is as follows: for each round $\bar{i} \leq i$, (i) top-$K$ settlement rates for round $\bar{i}$, (ii) the ordering between the order rates associated with the trader indexes $j \in [N_{\bar{i}}]$, (iii) for each $j \in [N_{\bar{i}}]$, the next round that this account participates in, (iv) aggregate marketplace fees for that round.

\begin{claim} \label{claim:orderrateconfidentiality}
\name{} provides relaxed order rate confidentiality.
\end{claim}
The simulator is provided the order book leakage and it sets the order rates of each trader in a way that matches the order book leakage. We describe how the simulator outputs are generated from the order book leakage while being indistinguishable to their real world counterparts.
During the order submission phase, information about an order rate is only used in the following ways:
(i) commitment on the order rate
(ii) the secret shares of the order rates to the broker nodes.
The hiding properties of the Pedersen commitments and the secret sharing protocol help in indistinguishability between the real and the simulated outputs.\\
After the order submission phase, information about order rates are used in the following entities that are revealed:
(i) the ordering between the order rates,
(ii) the top-$K$ settlement rates for the round, 
(iii) the aggregate of marketplace fees for that round, and
(iv) secret shares of the difference between the buyer and seller rates (as marketplace fees) to the broker nodes.
The first three are part of order book leakage input to the simulator and the simulator chooses the order rates to match with these outputs. The fourth entity can be easily generated in accordance with the aggregate marketplace fees.
% TODO: As discussed in sec:protocol, only sum of differences is revealed in every round
%Regarding (iii), the broker nodes only release the marketplace fees aggregated across trades in round and this can also be revealed through a private release. 
%but this gets revealed periodically and the system can be easily augmented to enable a private release.\\

\begin{claim}
\name{} provides front-running resilience.
\end{claim}
\name{} reveals no information on the order rates for the ongoing round before the matching phase begins, and this ensures front-running resilience.
The argument for this property is similar to the one in Claim \ref{claim:orderrateconfidentiality}, where the simulator obtains the order book leakage for all the previous rounds (and not for the ongoing round). The order rates for the ongoing round are chosen arbitrarily to match the order book leakage from the previous rounds. The hiding properties of the Pedersen commitments and the secret sharing protocol help in indistinguishability between the real and the simulated outputs for the ongoing round. And this proves that there is no leakage on the order rate (before the matching phase begins) since the simulator input here does not contain the order rate for the final round.

\begin{claim}
\name{} provides account balance confidentiality.
\end{claim}
%We simplify the argument with the assumption that all traders participate in the round when the account first participated in the marketplace. This assumption can be removed by using the 
We will argue that an adversary cannot differentiate between the two possible initial account balances if they admit the same order book leakage across rounds.
The hiding property ensures that the commitment scheme does not reveal information about account balance when a trader enters the marketplace. This along with the security of the secret sharing protocol and MPC protocols ensure account balance confidentiality.\\
The functions that involve account balance of a trader are (i) the comparison with the buy rate (for the buyer accounts) during submission, (ii) subtraction/addition of buy/sell rates during settlement. Hence, the information revealed about account is limited to the union of order book leakage for the rounds that a polynomial time adversary can associate this account to\footnote{The rounds of interest for an adversary to associate orders to this account is limited in a real world marketplace since the number of possibilities to keep track grows exponentially with the depth of the transaction graph rooted at the round where this account enters the marketplace, and this exponent has a large base with a large numbers of traders participating in each round.}.

\begin{claim}
\name{} provides trader unlinkability when account balance confidentiality and order rate confidentiality are provided.
\end{claim}
We will argue that the probability for any adversary to link a pair of orders across rounds $i$ and $i'$ to a trader is the maximum among $1/N_i$ and $1/N_{i'}$.
The order submission phase reveals the link between an account (i.e., commitment of account balance) that is participating in the round to the order it submits. But the oblivious shuffle protocol at the end of each round unlinks the account (old commitment) from its updated account (new commitment). Hence, the only information that is revealed after a round for all matched orders\footnote{Unmatched orders might continue to remain in the next round and hence the anonymity set is restricted to the traders whose orders get matched. Such a marketplace behaviour is not accounted for in this analysis.} is the subsequent round that each account (new commitment) in this round participates in. Any adversary is limited to a guess among the traders participating in the rounds, in addition to the negligible advantage in breaking account balance and order rate confidentiality and thus attempting to link traders through account balances. Hence it is the maximum of $1/N_i$ and $1/N_{i'}$ for rounds $i$ and $i'$.

\subsection{Market Integrity}
\label{sec:integrity}
We state the integrity of the \name{} marketplace.
\begin{claim}
\name{} ensures market integrity.
\end{claim}
At a high level, an adversary breaking market integrity with a non-negligible advantage breaks one of the following with a non-negligible advantage:
\begin{itemize}
\item soundness of the zero-knowledge proofs (order rate for a buyer is lesser than or equal to its account balance)
\item binding property of the commitment scheme (order rate used to generate the proof is the same as what is committed to)
\item correctness of the inputs used in MPC (semi-honest brokers are trusted to validate the input shares through their commitments on the blockchain; input share validation ensures this for malicious brokers)
\item correctness of the MPC protocol among the brokers
\item correctness of the smart contract execution (verify proofs and matching and settlement of orders as per the protocol)
\end{itemize}

\subsection{Quantifying Order Rate Privacy}
\label{sec:quantify}
In Section~\ref{sec:confidentiality}, we prove that the leakage of order rates is atmost \textit{order book leakage}.
% We analyse the privacy achieved by our protocol by quantifying the leakage of order rates.
  % We also quantify the dependence of privacy on certain key parameters, such as
  % the values for $K$ and $N$, the number of
  % top-$K$ settled orders revealed and the total number of orders in the round, respectively.

While an adversary (a trader or broker) cannot uncover any individual order placed by a specific trader, they can endeavor to estimate 
the distribution of orders in a given round based on the information leaked post-settlement.
We quantify the order book leakage to measure the privacy of the order rates.
% We introduce a notion of privacy, called 
% {\it Trade Confidentiality}, as a measure of privacy of order rates.
Specifically, we calculate the relative entropy of the estimated distribution of the order rates from its true distribution, in a trading round,
expressed as a percentage of the entropy of the true distribution. This measures how closely adversaries can guess the true distribution.
In Appendix~\ref{app:quantify}, we detail the analysis to measure the privacy gain (Equation~\ref{eqn:trade-confidentiality}).

\begin{align}
\text{Privacy Gain} &= \frac{\mathtt{KL(P_E, P_T)}}{\mathtt{H(P_T)}} \times 100
\label{eqn:trade-confidentiality}
\end{align}

\subsection{Discussion on Availability}
\input{discussion}

\input{comparison}

%%% Local Variables:
%%% mode: latex
%%% TeX-master: "paper"
%%% End:

%% file: discussion.tex
\label{sec:discussion}

% Property: Each round of the marketplace ensures maximal matching and fairness atleast to all the honest traders of that round.
We assume till now that (i) the brokers are available and (ii) blockchain network is available. These ensure that the marketplace is available.
% We discuss different scenarios in which the availability of the \name{} marketplace is maintained.
%We discuss some possible security attacks against the system by malicious traders or brokers, along with potential mitigations.

% \noindent  {\bf Front-running attacks:} Traders only share commitments of order prices on the blockchain, so no trader or broker
% can learn the actual orders placed, to launch a front-running attack. In the bucketization scheme, since the bucket ranges are
% revealed only in the second phase after receiving orders to be matched, traders do not learn even the number of orders in each
% bucket until after all orders for a round are placed. Traders learn the number of buy and sell orders that have been
% placed so far in an ongoing round, but this is legitimate information indicative of market supply-demand dynamics. 
% Front-running attacks on blockchain have been studied in~\cite{FrontRunning-FC19}.

% \textbf{Systems considerations:}
% The marketplace (i) excludes repeat orders from the same account during order submission and (ii) locks funds for each order placed before matching.

The rest of the section discusses this assumption on the availability of the blockchain network and the broker nodes.

\noindent {\it Flooding attack by malicious traders:} A malicious trader can launch a DoS attack by flooding 
the marketplace with a large number of orders that are unlikely to be matched.
The marketplace (i) excludes repeat orders from the same account during order submission and (ii) locks funds for each order placed before matching.
% The marketplace locks funds for each order placed, which acts as a natural deterrent for such attacks.
Further, the marketplace could charge a fee from traders for each submitted order, 
which can be refunded upon settlement. Unmatched orders are terminated after a predetermined number of rounds to refresh the order
book.

\noindent {\it Denial of service attacks on brokers:} A targeted attack on a single broker might restrict its participation in the MPC after receiving rate shares.
Note that this is covered with the attacked brokers being malicious.

\noindent {\it Denial of service attacks on blockchain:} The decentralized nature of blockchain is naturally resistant to DoS attacks. 
Having more blockchain peer nodes increases DoS resistance and is a well studied topic. In certain blockchain platforms, it may be possible
for a malicious peer or miner to delay or deny a trader from placing their order on blockchain. This requires care in designing the blockchain
platform to be resilient against malicious peers.

%% file: comparison.tex
\subsection{Comparison with Related Work}
\label{sec:comparison}

\begin{table*}
  \centering
  \captionsetup{justification=centering}
  \begin{tabular}{cccccccc}
    \hline

    Related & Account Balance & Order Rate & Trader & Front Running & Price & \multicolumn{2}{c}{Decentralization} \\
    \cline{7-8}

    Work & Confidentiality & Confidentiality & Anonymity & Resilience & Discovery & Matching & Settlement \\

    \hline

    % \multirow{2}{*}{ \makecell{Related Work}} & \multirow{2}{*}{\makecell{Account Balance \\ Confidentiality}} & \multirow{2}{*}{\makecell{Order Rate \\ Confidentiality}} & \multirow{2}{*}{\makecell{Trader \\ Anonymity}} & \multirow{2}{*}{\makecell{Front Running \\ Resilience}} & \multirow{2}{*}{\makecell{Price \\ Discovery}} & \multicolumn{2}{c}{\makecell{Decentralization}} \\

    % \cline{7-8}
    % & & & & & & Matching & Settlement \\
    % \hline

    FuturesMEX~\cite{FuturesMEX-SP18} & \cmark & \xmark & \cmark & \xmark & \cmark & \cmark & \cmark \\

    TEX~\cite{TEX} & \xmark & \xmark & \cmark & \cmark & \cmark & \xmark & \cmark \\

    ZEXE~\cite{ZEXE} & \xmark & \xmark & \xmark & \xmark & \xmark & \xmark & \cmark\\

    Dark MPC~\cite{darkMPC} & - & \cmark & \xmark & \cmark & \xmark & \cmark & - \\

    P2DEX~\cite{P2DEX} & \xmark & \xmark & \xmark & \cmark & \cmark & \cmark & \cmark  \\

    \name{} (Our Work) & \cmark & \hmark  & \cmark & \cmark & \cmark & \cmark & \cmark \\

    \hline
    \end{tabular}
    
    \caption{Comparison with related work.}
  \label{table:comparison}
\vspace{-0.1in}
\end{table*}

In Table~\ref{table:comparison}, we present a comparison of our work against relevant related work. {\cmark}/{\xmark} denote presence/absence of property in cited work. {\hmark} denotes the relaxed order rate confidentiality property of our work.

% Front-running resilience is ensured by the privacy of an order from other traders before the matching phase begins.
% Order rate confidentiality deals with privacy of order rates, even after matching and settlement are completed.
% While front-running resilience is supported by most marketplaces, order rate confidentiality is a more general privacy property to ensure only minimal necessary information about order rates is revealed, for price discovery.

FuturesMEX presents a distributed futures exchange where the order book is publicly visible, but trader anonymity is ensured.
It supports trader and inventory (account balance) anonymity, but does not provide order rate confidentiality or resilience to front-running attacks.
We do not consider futures contracts where trade positions are open for longer than a round of matching and settlement, and their attack 
scenarios are not directly applicable to our system that deals only with spot trades. It would nevertheless be a very interesting future work to combine 
the fully private order book in our system with futures contracts and margin buying supported by FuturesMEX.

TEX is a centralized marketplace protocol which supports front-running resilience through its time lock orders and receipts. However, the centralized operator learns the order rate after the trader reveals the secret key for matching.
% In TEX, the centralized operator is privy to the order rate only after the trader reveals the secret key.
% So, while it supports front running resilience, it does not achieve order rate confidentiality against the operator.
Also, while the settlement is non-custodial and decentralized, by verification using ZKPs, account balances are revealed to the operator.
% TODO: is their settlement privacy preserving or not?

ZEXE introduces a protocol for private, secure off-chain computation by an operator with public on-chain verifiable transactions. Their closed-book DEX supports order rate confidentiality, front-running resilience and trader anonymity against traders but not against the non-custodial order book operator.
% TODO: Is the order book operator centralized/decentralized - they have not mentioned in their paper?
% TODO: Is settlement privacy preserving or not?
The settlement and account balances are private to traders but not to the order book operator.
Since order rates are private to traders, they do not support price discovery.

Dark MPC presents a privacy-preserving matching protocol for CDA (Continuous Double Auction) in dark markets using MPC. However, they do not support ownership and settlement of assets (represented by `-' in Table~\ref{table:comparison} under account balance privacy and decentralized settlement). Also, their protocol does not support price discovery. 

A concurrent work, P2DEX presents a privacy-preserving decentralized exchange using publicly verifiable MPC to perform privacy-preserving matching on orders.
However, settlement transactions, using the UTXO model, are on plaintext order prices, thereby not supporting order rate confidentiality.
Prior work ~\cite{ron2013quantitative, reid2013analysis} has shown that transaction tracing and transaction graph analysis results in de-anonymization of Bitcoin, affecting both anonymity and account balance privacy.

In Appendix~\ref{appendix:AMM}, we discuss about Automated Market Maker and briefly describe the design of a privacy-preserving AMM DEX.

% Our work deals with limit orders in a call market, where orders are collected and executed at predetermined time intervals, as opposed to continuous markets where traders continuously submit orders and which are continuously executed against an order book of standing orders.
% A market maker is an entity in a marketplace which quotes both buy and sell orders of an asset in order to make a profit.

%%% Local Variables:
%%% mode: latex
%%% TeX-master: "paper"
%%% End:

%% file: evaluation.tex
\label{sec:evaluation}

In this section, we present the details of the implementation and the performance evaluation of the \name{} protocol.

\subsection{Implementation}
\input{implementation}

\subsection{Evaluation}
We present a detailed evaluation of the performance of the
% \textsc{Bucketization} and
\name{}
protocol and use the alternatives mentioned in Section~\ref{sec:overview} as baselines for comparison.
Our goal is to demonstrate that we can achieve high levels of privacy and decentralization with acceptable trade-offs on performance.
All experiments are performed on a 14-core 64-bit Intel Xeon Gold 6132 CPU @ 2.60GHz. 
%All traders are pre-registered with the blockchain with a unique identity represented by a digital certificate. 
The exchange smart contract is initialized with an account for each trader with a certain balance (or a commitment of the balance
for our privacy-preserving schemes) and ownership of
a certain quantity of the asset to be traded in the marketplace.

During each experiment, trader processes continuously submit buy and sell orders to the marketplace following a poisson process
with a certain mean. Their order price
is an integer drawn from a uniform distribution with a mean value of 255 for buyers and 245 for sellers and a variance of 15 (the distribution
of prices does not impact performance).
Periodically, matching is triggered
on all outstanding trades submitted, followed by settlement. We refer to the completion of each instance of matching and settlement
as a round. Each experiment lasts for $12$ rounds and each data point in our plots is obtained by averaging across $12$ experiments.
Unless specified otherwise, we use the following default values: 512 orders submitted on average in each round, 
3 brokers,
 round time of 30 seconds (that is, matching is triggered every 30 seconds) and a marketplace
comprising of 4 blockchain peers. For the \textsc{Bucketization} scheme, we use a default bucket width of 4.
We study the performance of our protocols by varying each of these
parameters.  
%% Done: Fill X and Y values above

We measure the following metrics:
\begin{itemize}
\setlength\itemsep{0em}
 \item \textit{Average end-to-end latency:} This is the primary metric essential to a trader and is the average time taken from 
submission of an order until it is settled.

 \item \textit{Component-wise latency:} The end-to-end latency comprises of waiting time until the next matching round, matching time,
settlement time and for \name{}, the time taken for sorting using MPC. 
%This is useful in understanding
%trade-offs at different components in the exchange.

 \item \textit{Percentage of orders matched}

 \item \textit{Marketplace fees} as a $\%$ of total worth of trades settled  
\end{itemize}

\begin{figure}[]
    \centering
      \includegraphics[width=1.0\columnwidth]{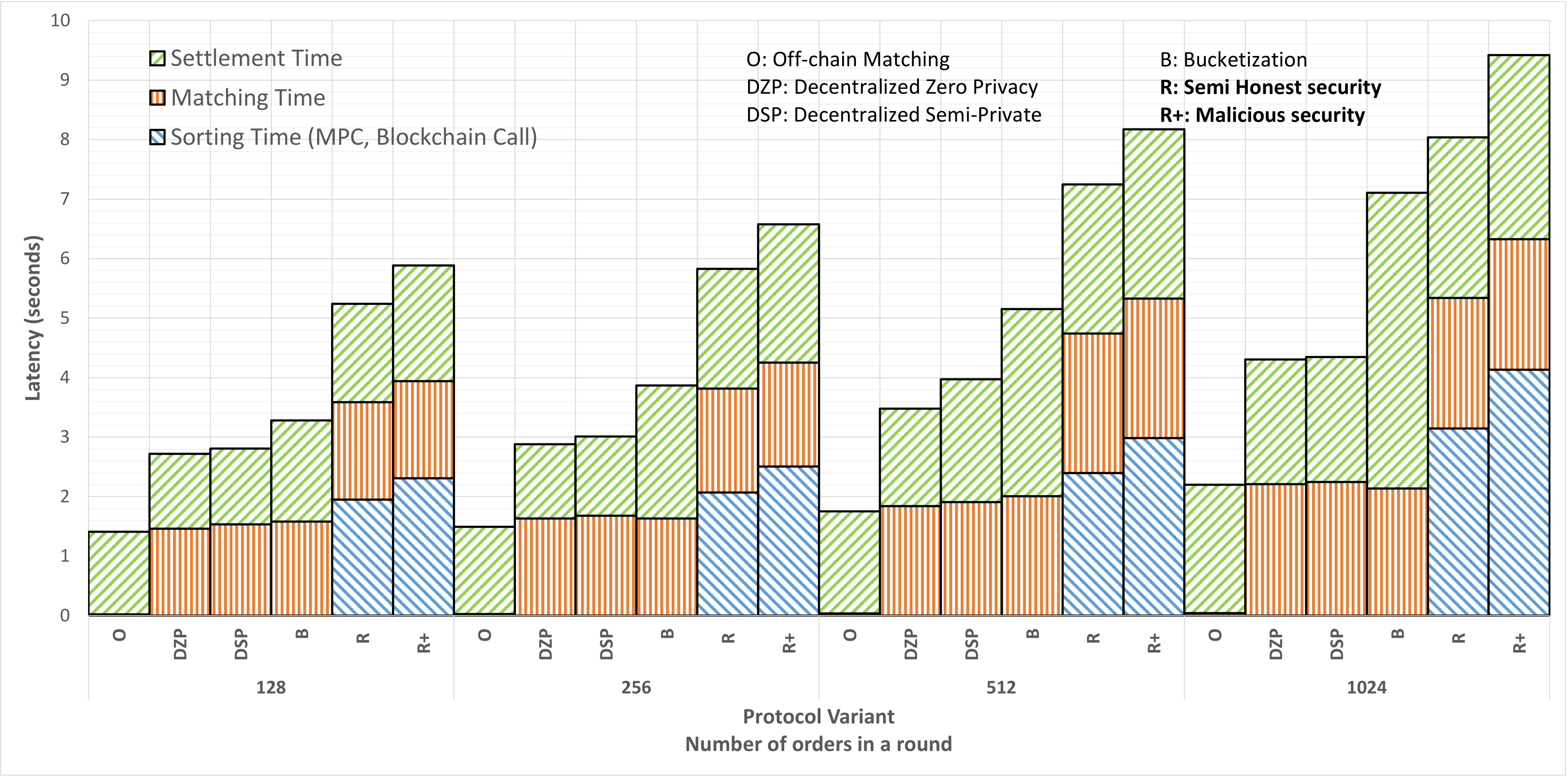}
      \captionsetup{justification=centering}
    \caption{Component-wise and end-to-end latency comparison of all 6 protocols for increasing order volume}
    \label{graph:latency-orders}
\end{figure}

\noindent \textbf{Scaling number of orders:}
In this experiment, we compare the component-wise and end-to-end latency of all the proposed protocol variants for increasing number of orders in
each round and present the results in Figure~\ref{graph:latency-orders}. 
%Per commodity trading volumes in some of the largest centralized commodity exchanges range from few tens to few hundreds of trades a minute.
Centralized commodity exchanges today handle up to a few hundred orders a minute per commodity.
To demonstrate that our protocols can scale to such volumes, we present results for 
128, 256, 512 and 1024 orders generated in each round comprising of 30 seconds. The non-privacy preserving, but decentralized protocol variants incur hardly any 
additional latency with increasing number of orders, showing that the use of blockchain smart contracts scales very well with load.
The \textsc{Bucketization} and \name{} protocols exhibit a marginal sub-linear increase in latency with increasing load,
significantly surpassing the scaling needs of most commodities traded in exchange markets today. For handling a large number of commodities
in parallel, the underlying infrastructure can be suitably scaled.

The \textsc{Centralized} exchange (not shown in the graph) without any privacy or decentralization takes negligible time for matching and settlement and is the most performant. 
For the \textsc{Offchain Matching} protocol, 
matching is performed by a centralized entity outside blockchain and the main latency incurred is the cost of performing settlement on blockchain.
The \textsc{Decentralized Zero Privacy} protocol %also performs matching on the blockchain, so 
incurs a small added delay due to the use of blockchain
for both matching and settlement. The \textsc{Decentralized Semi Private} protocol, additionally leverages Pedersen commitments to store the
account balances in private and hence incurs slightly higher settlement times.
% compared to \textsc{Decentralized Zero Privacy} protocol. 
For the \textsc{Bucketization} protocol, since matching is based on public bucket floor and ceiling values, it is as performant as the previous protocols.
However, settlement based on the private order rates of the traders incurs additional overhead for communication between the matched traders, recording
transactions on blockchain and for zero knowledge proofs. As expected, this overhead during settlement is also higher for larger volume of orders, 
but scaled sub-linearly as matched pairs of buyers and sellers can execute the settlement process in parallel. 
%We believe there is substantial scope for
%parallellization in our implementation that could further reduce the overheads incurred by our privacy-preserving protocols.
Finally, the \name{} protocols experience the highest latencies for matching and settlement largely owing to the overhead of
MPC. The \SortingMPC\ time, although performed by a fixed set of 3 brokers, increases nearly linearly with the number of orders due to the 
complexity of sorting.
With the duration of a round set to 30 seconds,
% and orders generated uniformly following a poisson process, 
all trade orders will have a waiting time (not shown in graph) of an average of 15 seconds for all the protocols. The marketplace is mostly idle during this time, except for verifying proofs.
\begin{figure}[!t]
%  \begin{minipage}[t]{0.5\textwidth}
    \centering
      \includegraphics[width=0.9\linewidth]{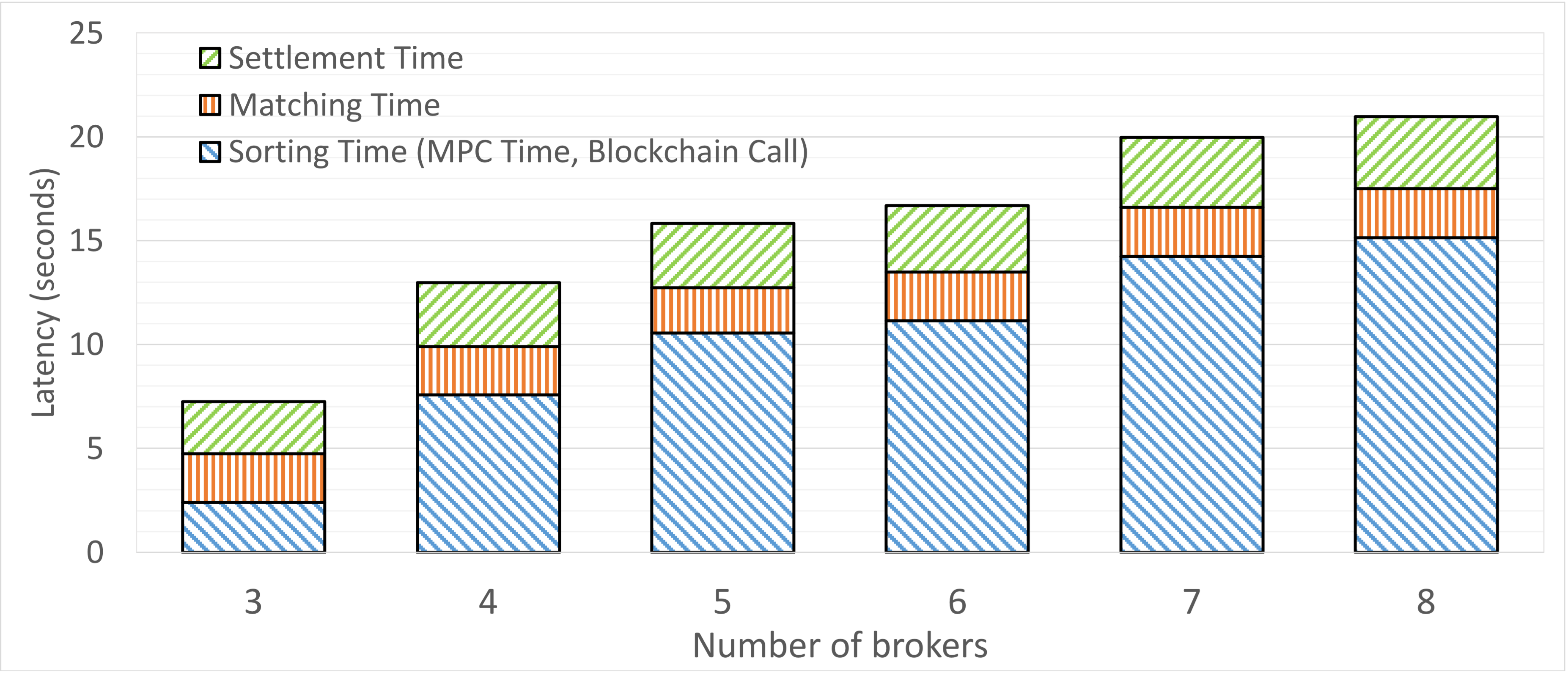}
    \caption{Latency in \name{} for different number of brokers, for semi-honest security.}
    \label{graph:brokers}

\end{figure}

\noindent \textbf{Scaling number of brokers:}
The choice of number of brokers presents a crucial trade-off between performance and the level of tolerance to collusion among the brokers.
Privacy is lost if all brokers collude to compute the secret order price of traders.
Figure~\ref{graph:brokers} presents a comparison of the component-wise latency for the \name{} protocol with semi-honest security 
when the number of brokers is varied from 3 to 8 with 512 orders submitted per round.
With an increase in the number of brokers, the time taken by the MPC for sorting orders increases linearly, while matching and settlement times
remain nearly the same.

\begin{figure}[t!]
    \centering
    \includegraphics[width=0.9\linewidth]{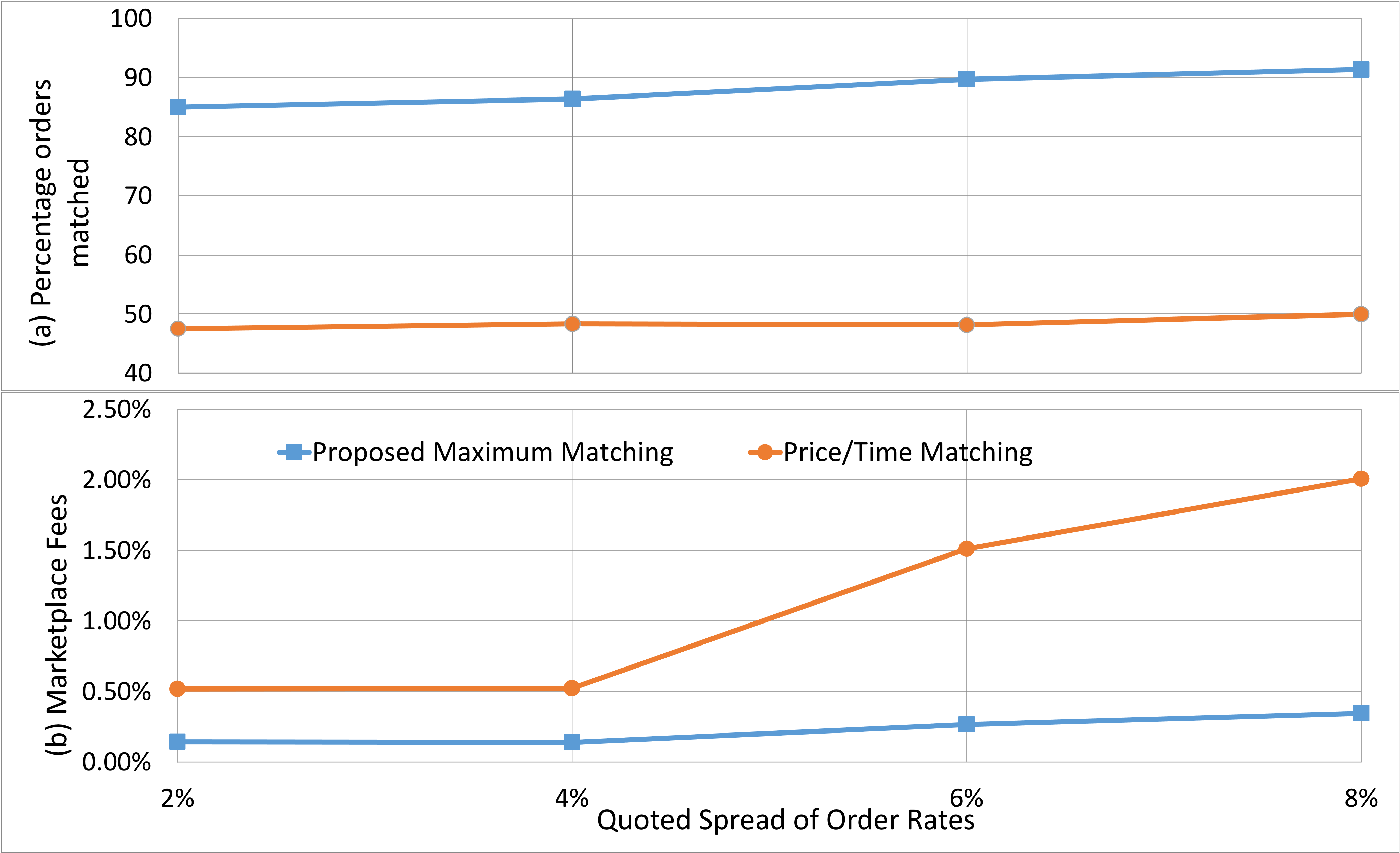}
    \caption{Percentage of orders matched and marketplace fees when quoted spread of order rates is varied.}
    \vspace{-0.1in}
    \label{graph:bidask}

\end{figure}
\noindent \textbf{Varying bid-ask spread of order rates:}
Bid-ask spread metrics~\cite{amihud1986asset} 
are used as a measure of liquidity of the marketplace. The quoted spread is defined as the difference between the lowest asking price 
by sellers and the highest bid price by buyers, as a fraction of their mean. We use different values of the quoted spread to
determine the range of bid and ask prices around the mean.
We have varied the bid-ask quoted spread, from 2\% to 8\% while keeping the mean the same for both buy and sell orders to generate order rates using a uniform distribution.
This has no effect on the observed latencies. However, the spread would affect the percentage of orders that are matched and the fees
that the marketplace can earn. For this experiment, we study the scenario where the marketplace earns the difference between matched buyer and
seller prices. We compare our maximal and fair matching algorithm with the price-time matching algorithm.
Figure~\ref{graph:bidask}a depicts the percentage of orders matched for the two matching algorithms. % for different bid-ask quoted spreads.
As expected, the price-time matching algorithm achieves a matching percentage of about 50\%, whereas our maximal matching algorithm matches 85-90\% of orders.
In contrast, the price-time algorithm allows for significantly higher marketplace fees as shown in Figure~\ref{graph:bidask}b. This is because price-time
matches the most competitive buyer to the most competitive seller that matches its price, leading to greater disparity between buyer and seller rates. It
does not attempt to maximize the number of matches, which would reduce the difference between matched buyer and seller rates.
% ToDo:
We also ran the same experiment for normal distribution of order rates and observed similar results.

\noindent\textbf{Blockchain scalability}:
We varied the number of blockchain peer nodes from 4 to 12, which did not impact the latencies for the \name{} protocol (plot not shown). 
Our blockchain based marketplace scales quite well to a large number of blockchain nodes.

\noindent \textbf{Communication Overheads:}
The communication costs in the system occur due to the blockchain and MPC components. 
%As shown in previous experiments, the
Blockchain overheads are quite low with our use of Hyperledger Fabric, with blocks having a size of at most a few hundred kilobytes. 
% ToDo: add Fabric costs
For the default parameters 
of $512$ orders in a round and $3$ broker nodes, each party communicates $1.38$ MB for MPC for semi-honest security and $20.79$ MB for malicious security.
This overhead is higher for a larger number of brokers, but is still easily achievable even across wide area networks.

\begin{figure}[t]
    \centering
      \includegraphics[width=0.9\linewidth]{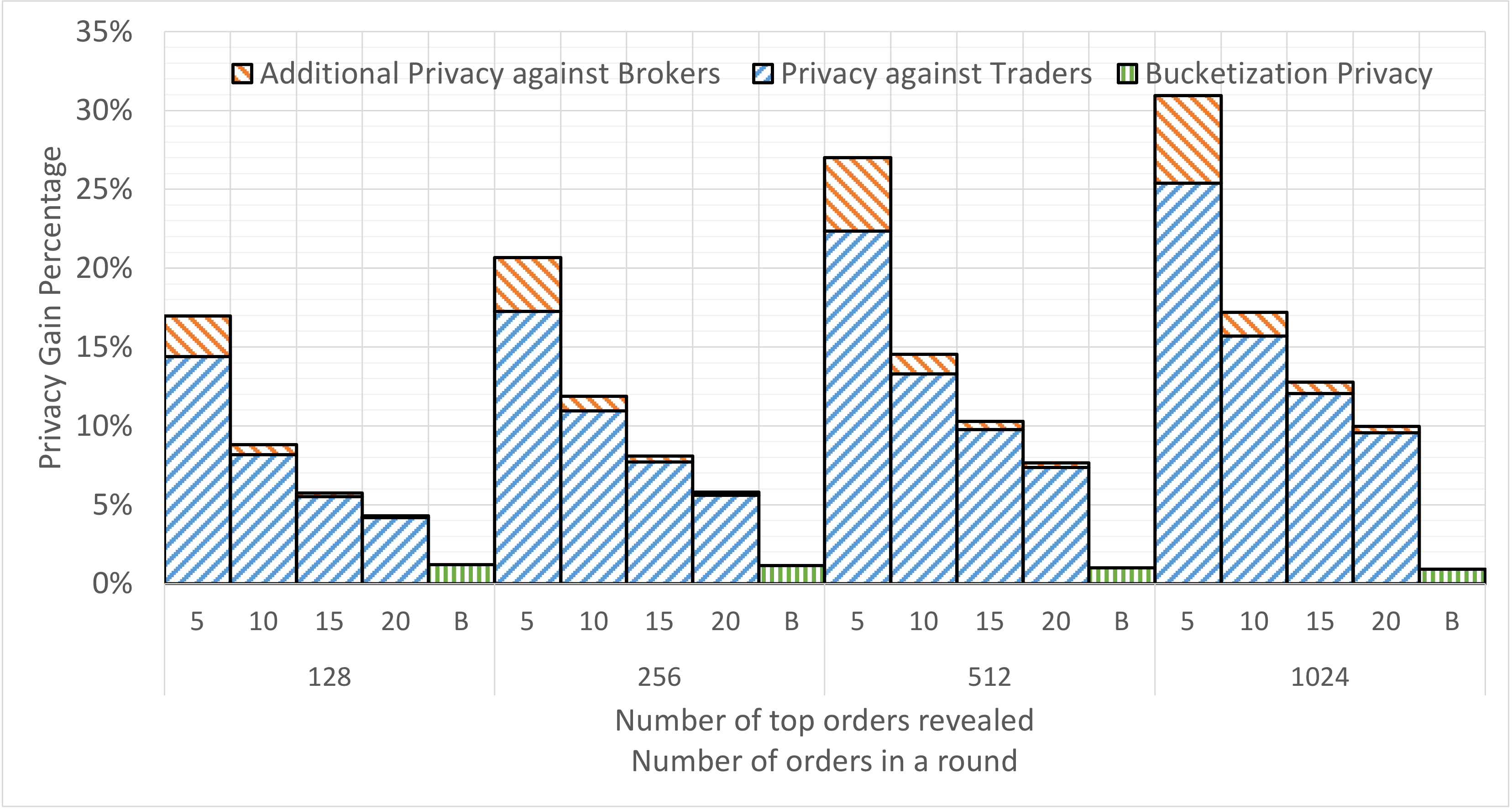}
    \caption{Privacy of \name{} for traders and brokers for increasing order volume and number of top matched orders revealed.}
    \label{graph:mprivacy}
    % \end{minipage}
\end{figure}

\noindent \textbf{Privacy Measure:}
As discussed in Section~\ref{sec:quantify}, 
we defined the privacy gain of the marketplace as the KL divergence of the estimated distribution from the true distribution as a percentage of the entropy of the true distribution.

For order rates sampled from a normal distribution and averaged across 100 runs, we computed the privacy gain of the \name{}
and the \textsc{Bucketization} protocols
for varying number of top matched orders revealed ($K$) and the number of orders ($N$) in each round. The results are presented in Figure~\ref{graph:mprivacy}.
For the \name{} protocol, the marketplace has higher privacy against brokers than against traders, who have higher information gain from their own order rate.
As expected, privacy gain is higher for lower $K$, as lesser information is revealed. For a fixed $K$, privacy gain increases with the number of orders in a round as many more values are hidden.
The privacy of the \textsc{Bucketization} protocol does not vary much with $K$ since adversaries can approximate other unknown order rates to the granularity of a bucket.
% Privacy decreases with the number of orders, since a larger sample allows estimation of population parameters with greater accuracy.
% Privacy increases with bucket width, increasing from $0.2\%$ for width $2$ to $1.2\%$ for width $16$ (not shown as values are low). 
%%% Local Variables:
%%% mode: latex
%%% TeX-master: "paper"
%%% End:

%% file: implementation.tex
\label{sec:implementation}
%We have implemented the centralized and decentralized marketplace protocols introduced and described in \ref{sec:overview} and \ref{sec:protocol}.

\textbf{Blockchain Platform:}
We implemented the decentralized blockchain based marketplace protocols using Hyperledger Fabric~\cite{Fabric-Eurosys19}.
It is a permissioned blockchain platform which, as opposed to permissionless blockchains, leverages strong identities based
on digital certificates to enable permissioned membership to the blockchain for both peer nodes as well 
as clients. It has support for smart contracts with the ability to restrict access
to functions and data to specified roles and identities. 
Traders are implemented as clients with unique identities on the blockchain.
Marketplace functionalities of managing account balances for traders, maintaining an order book, and performing matching and settlement
are implemented as smart contracts (\textit{chaincodes}) on Hyperledger Fabric, written in Golang. 

\noindent \textbf{MPC:}
Brokers in \name{} are implemented using the open source MP-SPDZ framework~\cite{MPSPDZ, keller2020mp}.
% which provides a suite of implementations of MPC protocols supporting multiple parties.
For the semi-honest protocol, we use the replicated-ring-party mode for 3 parties and shamir-party mode for more than 3 parties.
For the malicious version, we use the ps-rep-ring-party mode.

\noindent \textbf{Range proofs and Pedersen Commitments:}
All the proposed protocols require ZK proofs of range for proof of account balance.
For the ZK proofs of range, we have used bulletproofs~\cite{bunz2018bulletproofs}, which is a non-interactive ZK proof protocol which supports very efficient range proofs. 
We have used an open source Rust implementation of bulletproofs~\cite{Bulletproofs} which also provides support for Pedersen Commitments on the Edwards25519 elliptic curve.
Since the proofs are verified by the marketplace facilitators on blockchain, we have written a Rust wrapper around the bulletproofs package which is compiled as a static library 
and used with the smart contracts in golang using Cgo.

%%% Local Variables:
%%% mode: latex
%%% TeX-master: "paper"
%%% End:

%% file: conclusion.tex
\label{sec:conclusion}
\vspace{-0.05in}
In this paper, we present the design of \name, a fully decentralized privacy-preserving exchange marketplace
with matching, automated on-chain settlement and price discovery
supporting marketplace properties such as order rate and account balance confidentiality, unlinkability between trade orders and traders, front-running resilience and fairness.
% using blockchain and cryptographic tools like zero knowledge proofs and multiparty computation.
% We have proposed a matching algorithm to ensure maximality of matching and fairness to buyers and sellers.
% Further, we have presented a range of protocols showing possible trade-offs between privacy and performance. 
We define formal security notions for the marketplace and present a security analysis of our protocol.
We quantify information leakage to allow the marketplace to determine parameters in accordance with its privacy needs.
We demonstrate that our proposed solutions scale well and are suitable for real world markets for a large class of commodities.
% We have also discussed how our system is resilient to security attacks such as front-running. 

As described in our paper, marketplaces need to occasionally reveal the marketplace statistics to help future participants in discovering the current market price for an asset.
However, this affects order rate confidentiality, presenting a trade-off between confidentiality and utility.
One interesting future work would be to calculate and reveal such marketplace statistics in a differentially private manner without affecting the marketplace properties like market integrity.
%%% Local Variables:
%%% mode: latex
%%% TeX-master: "paper"
%%% End:

%% file: appendix.tex
\section{Bucketization Based Protocol}
\label{sec:appendix-bucketization}

In Section~\ref{app:bucket-protocol}, we describe the order submission, matching and settlement for our bucketization protocol. In Section~\ref{app:bucket-privacy} we describe the privacy quantification for this scheme as evaluated in Section~\ref{sec:evaluation}.

\subsection{Protocol}
\label{app:bucket-protocol}
\textbf{Order submission:} If the bucket start and end values were known apriori to the traders, a buyer would always choose 
the lower end of their intended bucket and a seller would choose the higher end, with no effect on the matching.
To keep the order rates dynamic, we divide the order submission into two phases. In the first phase, traders submit
orders with rate commitments.
Buyers additionally submit a ZKP of range that their order rate is less than or equal to their account balance, verified in the smart contract.
After T seconds or receiving M orders, whichever is earlier, the marketplace chooses random bucket 
start and end values for this round and publishes this on the blockchain.
Deterministic random numbers can be generated within smart contracts using the block hash or timestamp as a seed~\cite{kraft2019game} or using Verifiable Random Functions~\cite{algorand,praos}.
% Exact protocol:
The smart contract uses the hash of the current block as a seed to generate a deterministic random number while ensuring that traders cannot predict them.
% The histogram of orders in buckets in the previous round can also be used together with randomization in generating the new bucket ranges.
Then in the second phase, within a timeout of T' seconds, 
all traders must submit the bucket that their rate belongs to, along with a zero knowledge proof of range.
For buy orders, the blockchain locks funds equalling the buyer's rate in an escrow using the commitment of the rate.

\textbf{Matching:} The marketplace performs matching using the bucket values of the orders. % using Algorithm~\ref{algo:matching}. 
In case of buyers, we use the floor value of the bucket that their order price belongs to, and for sellers, we use the bucket ceiling value during matching. 
This ensures that any matching of counterparties that we find using our matching algorithm will definitely be feasible with respect to their private
rates as well, as we are considering stricter constraints.
% Buy orders are sorted in non-decreasing order of their bucket floor rates and sell orders are sorted in non-decreasing order of their bucket ceiling rates.
The input to the algorithm is the set of orders sorted in increasing order of bucket values, with ties broken by timestamp, where an earlier submitted order
%with an earlier timestamp on the blockchain
gets precedence.
% Each buyer is matched with the first seller whose rate is lower than the buyer's rate. 
Matching is performed similar to the \name{} protocol.
Note that the matching algorithm matches a buyer in a higher bucket with a seller in a lower bucket and will not match orders 
within a bucket. This is because a buy order might have lower order rate than a sell order within the same bucket, which is an invalid match.
We claim that this matching algorithm which considers buyers and sellers in increasing bucket values produces a maximum matching, 
among all matching algorithms with buckets and prove the same in Appendix~\ref{appendix:matching-proof}.

\textbf{Settlement:} We propose two schemes for computation of settlement rate and execution of settlement between
each matched pair of buyer and seller using
the traders' private proposed rates. Each matched buyer and seller is required to communicate their order rates to their
counterparty, so that they are able to compute their individual final account balance after settlement.

\textit{Marketplace earns the rate difference:} In this scheme, each trader settles at their respective proposed rates and the difference between the buyer 
and seller rates for each matched pair is paid as fees to the marketplace.
The protocol proceeds as follows. In the first step, each buyer and seller encrypt their proposed rate and blinding factor used in the commitment of their
order along with a digital signature,
with the public key of their matched counterparty and record it on the blockchain. This helps the marketplace verify fairness of the exchange, so 
no party can claim that it has not received a message from the other party. Each trader decrypts their message, verifies the signature and verifies that 
the values match the commitment submitted by their counterparty on the blockchain. In case the values do not match, they notify the marketplace 
of the deviation using the digital signature as proof and the cheating party is penalized.
In the second step, each trader calculates the marketplace fees as the difference of the rates and submits it to the blockchain along with a ZKP that the difference 
of the commitments of the buyer and seller rates on the blockchain opens to the submitted value.
% (the commitments are homomorphic).
% This is a ZKPoK of discrete logarithm.
The marketplace verifies the proof with a smart contract on the blockchain
and settles the traders by adding the seller's commitment to the seller's account balance and 
adding the difference value to its own account (potentially shared by the entities operating the peers of the decentralized marketplace) and removing the funds held in the escrow on behalf of the buyer.

\textit{Settlement at mean of traders' rates:} In this scheme, the traders settle at the mean of their rates (the marketplace can charge a fixed fee or a percentage of 
the settlement rate as fee). The first step of the traders' securely exchanging their proposed rates and the blinding factors with their counterparty is the same as above.

The smart contract calculates settlement rate as the mean of commitments of buyer and seller (since PC are homomorphic), removes the funds locked in the escrow on behalf of the buyer, returns the excess funds 
(difference between proposed rate and settlement rate) to the buyer's account and adds the settlement funds to the seller's account.

\section{Quantifying Order Rate Privacy}

\subsection{Rialto}
\label{app:quantify}

In this section, we present an analysis of the privacy achieved by \name{}, by quantifying the leakage of order rates.
  We also quantify the dependence of privacy on certain key parameters, such as
  the values for $K$ and $N$, the number of
  top-$K$ settled orders revealed and the total number of orders in the round, respectively.

% We have introduced  {\it Trade Confidentiality} to quantify privacy by measuring how closely adversaries can guess the true distribution.
% It is calculated as the relative entropy of an adversary's estimated distribution from the true distribution of order rates in a trading round,
% as a percentage of the entropy of the true distribution.

For our analysis, we assume that the true distribution of order rates of all buyers and sellers follows a single Gaussian distribution $\mathtt{P_T}$ 
with mean $\mu_T$ and variance $\sigma_T^2$, unknown to all the participants. While real world order rates 
might have different distributions for buyer and seller rates, we make this simplifying assumption to make privacy measures uniform across all traders.
We also assume that traders submit only one order in a round and knowledge is not carried across rounds of the protocol.

The marketplace reveals the top-$K$ order rates for price discovery along with a 
total sorting of the orders as output by the \SortingMPC. Adverseries estimate Gaussian parameters, $\mu_E$ and $\sigma_E$, from this information.

Blom~\cite{Blom} proposes the following equation for the expectation of the $r^{th}$ order statistic of a Gaussian distribution,
where $\alpha = \frac{\pi}{8}$ is a constant and 
$\Phi^{-1}$ is the quantile function of the standard Gaussian distribution.

\begin{align*}
\text{Blom}(r, n) &= \mu_E - \Phi^{-1}\left(\frac{r - \alpha}{n - 2\alpha + 2}\right).\sigma_E \\ 
\end{align*}

At the completion of any trading round, brokers can use Blom's equation using the top-$K$ matched order rates, which are the top $K$ order statistics, 
to numerically solve for parameter estimates $\mu_E$ and $\sigma_E$.
Traders can additionally use their own order rate, whose $r$ is revealed from the ranking of orders, to get more accurate 
estimates $\mu_E$ and $\sigma_E$.
In our evaluation in Section~\ref{sec:evaluation}, for orders generated from a true Gaussian distribution with parameters
$\mu_T$ and $\sigma_T$, we use the above analysis to obtain estimates $\mu_E$ and $\sigma_E$.
Finally, we measure the privacy gain (Equation~\ref{eqn:trade-confidentiality}),
as the KL divergence between the Gaussian with estimated parameters ($P_E$) and the true Gaussian ($P_T$) as a percentage of the entropy of the true distribution ($H(P_T)$).

% \vspace{-0.1in}
% \begin{align}
% \text{Privacy Gain} &= \frac{\mathtt{KL(P_E, P_T)}}{\mathtt{H(P_T)}} \times 100
% \label{eqn:trade-confidentiality}
% \end{align}

\subsection{Bucketization Protocol}
\label{app:bucket-privacy}

In the \textsc{Bucketization} protocol, the marketplace leaks a histogram of orders across buckets.
Adversaries extrapolate the true distribution parameters from the histogram statistics and the top-$K$ settlement rates revealed for price discovery.
Privacy gain will be a function of the bucket width $W$ and $N$, the expected number of orders in a round, allowing the marketplace to choose parameters depending on its privacy requirements.

Adversaries estimate each of the $N$ orders rates in the round by sampling randomly, assuming (say) a uniform distribution within each bucket and using exact values for the top-$K$ order rates and their own rate.
They can then estimate the population (true Gaussian distribution)'s mean as the sample's mean and the population's variance by scaling the sample's variance by $N$ to account for the smaller variance of a smaller sample.

We measure the privacy gain of the marketplace by computing the Kullback-Leibler divergence ($\mathtt{KL}$ divergence) of the 
estimated Gaussian from the true Gaussian distribution, expressed as a percentage of the entropy of the true distribution ($\mathtt{H(P_T)}$) as per Equation~\ref{eqn:trade-confidentiality}.

We find that privacy decreases with the number of orders, since a larger sample allows estimation of population parameters with greater accuracy.
Privacy also increases with bucket width, increasing from $0.2\%$ for width $2$ to $1.2\%$ for width $16$. 

\section{Automated Market Maker}
\label{appendix:AMM}
Automated Market Maker (AMM) such as~\cite{berg2009hanson} is an automated system or algorithm to settle traders that uses a liquidity pool instead of a limit-order book for matching.
Recent work~\cite{xu2021sok,capponi2021adoption} show a rising adoption of AMM based Decentralized Exchange (DEX) for DeFi (Decentralized Finance) blockchain applications.
An AMM based DEX exchanging two tokens has a liquidity pool for the token pair allowing direct exchange between the two tokens.
An AMM uses a conservation function to determine the exchange rate of the asset depending on the quantity of the token assets,
such as a constant product AMM where the product of the token assets is constant.
% A constant product AMM with a liquidity pool with $T_X$ tokens of $X$ and $T_Y$ tokens of $Y$, operates such that $T_x.T_Y$ is a constant. If a trader wishes to exchange $\delta_X$ tokens of $X$, it receives $\delta_Y$ tokens of $Y$ such that $T_X.T_Y = (T_X + \delta_X).(T_Y - \delta_Y)$.

While the design and analysis of a privacy-preserving AMM DEX could be an interesting future work, we briefly describe the design of an AMM dealing with assets $X$ and $Y$ with token amounts $T_X$ and $T_Y$ respectively, which preserves the privacy of the order rate and the exchange rate.

\begin{enumerate}
\item {\bf Initialization:} The liquidity pool is initialized by Liquidity Providers which provide tokens and receive pool shares in proportion to their contribution.
Liquidity providers lock their tokens in an escrow and secret share their contribution among the brokers, which verify their shares.
The total pool liquidity itself is hidden, maintained as secret shared between the brokers.

\item {\bf Trade}: A trader submits an order to buy/sell tokens of $Y$. We can also support a trader to set a limit for the maximum/minimum tokens of $X$ they are willing to exchange (in private, similar to \name).
The marketplace calculates the exchange rate using MPC and checks if it matches the limits specified by the trader.

\item {\bf Settlement:} The marketplace creates a settlement transaction to the trader, transferring the exchanged tokens in a privacy-preserving manner using PC.

\item {\bf Price Discovery:} Finally, the marketplace could periodically reveal the pool liquidity to help with price discovery.

\end{enumerate}

\section{Proof of Maximal Matching}
\label{appendix:matching-proof}
We prove maximality of our matching algorithm using the graph theory definition of matching. We model the buyers and sellers as nodes of a bipartite graph $G$.
There exists an edge between a buyer and seller, if the buyer's rate is no less than the seller's rate. The result of any graph matching algorithm
on this bipartite graph represents a feasible matching.

% The edge condition between traders $T_1$ and $T_2$ is as follows:
% $T_1 > T_2$ iff $T_1$ occurs after $T_2$ in the ranking of orders calculated by the brokers.
An edge occurs between traders $T_1$ and $T_2$ if $T_1 > T_2$ ($T_1$ occurs after $T_2$ in the sorted list of orders).

A matching $M$ is maximum if there does not exist any augmenting path. An augmenting path is a path between two distinct unmatched vertices 
where the edges are alternately in $M$ and not in $M$.
We claim that our matching ($M_1$) is maximum. 

Suppose by contradiction, it is not a maximum matching. Then, there must exist an augmenting path consisting of at least $3$ edges.
Let this augmenting path be $B_1$ - $S_1$ - $B_2$ - $S_2$, where $S_1$ and $B_2$ are matched with each other in $M_1$, while $B_1$ and $S_2$ are both unmatched. 
Inverting the augmenting path, where $B_1$ is matched with $S_1$ and $B_2$ is matched with $S_2$, will produce a matching $M_2$
of greater cardinality.
%with cardinality larger than that of $M_1$.

We list the following facts about our matching $M_1$:
\begin{enumerate}
\setlength\itemsep{0em}
\item It must be that $B_1 < S_2$. Otherwise, our matching would have matched $B_1$ with $S_2$ and they would not both be unmatched.
\item Since $B_2$ and not $B_1$ is matched with $S_1$, it must be that $B_2 \leq B_1$. Otherwise, $B_1$ would have been considered earlier for matching by the algorithm and would be matched.
\end{enumerate}

This implies that $B_2 < S_2$. This is a contradiction as then, an edge cannot exist between $B_2$ and $S_2$ by the construction of the graph, 
for it to be part of the augmenting path and matching $M_2$. 
Hence, no such larger matching $M_2$ can exist. Thus, our matching is a maximum matching.

\section{Algorithms}
\label{sec:appendix-algorithms}

\algblockdefx[DOP]{Do}{EndDo}[0]
{\textbf{do in background}}
{\algorithmicend}

\newcommand{\Wait}[2]{\State \textbf{Wait for}\ #1; \algorithmicthen\ #2}
\newcommand{\IFor}[2]{[#1\ \algorithmicfor\ #2]}
\newcommand{\IForEnd}{\unskip}
\newcommand{\RecvMsg}[2]{#1 <- \textsc{\hl{#2}}}
\newcommand{\SendMsg}[2]{\textsc{\hl{#1}}(#2)}
\definecolor{light-gray}{gray}{0.9} 

%%%%
% Order Submission
% 1. ReceiveOrder(commitment)
% 2. Buckets()
% 3. ReceiveBucket(bucket, proof):
%         verify 
%
% Matching
% 1. matching()
%
% Settlement
% 1. SubmitRate()
%     digital signature
%     encrypt
%     write to ledger
% 2. SubmitSettlementRate()
%     commitment to settlement rate / difference + ZKP
%

%%%%
% Order Submission
% 1. ReceiveRateShare(share, commitment)
%
% Matching
% 1. PerformMPC()
% 2. Matching()
%
% Settlement
% 1. SettlementMPC()
% 2. Settle()

We list the algorithms used in the paper.
Algorithm~\ref{algo:fair-swapping} is the swapping algorithm to swap unmatched buy orders with matched buy orders for fairness as discussed in Section~\ref{sec:fairness}.
Algorithm~\ref{algo:matching} is the matching algorithm used by the marketplace to match buyers and sellers.
Algorithms~\ref{algo:Mmarket},~\ref{algo:Mbroker} and \ref{algo:Mtrader} show the sequence of steps at the marketplace, a trader and a broker respectively for the \name{} protocol.
Algorithms~\ref{algo:bmarket} and \ref{algo:btrader} show the sequence of steps for the marketplace and a trader respectively for the \textsc{Bucketization protocol}.

\begin{algorithm}
\caption{Swapping algorithm for fairness}
\label{algo:fair-swapping}
\begin{algorithmic}[1]
\fontsize{10pt}{10pt}\selectfont
\Procedure{swapping}{MatchedOrders, SortedOrders}%\Comment{}
\While {Exists unmatched buyer with higher price than a matched buyer}
    \State swap highest price unmatched buyer replacing highest price matched buyer lesser than it
\EndWhile

\EndProcedure
\end{algorithmic}
\end{algorithm}

\begin{algorithm}
\caption{Matching algorithm}
\label{algo:matching}
\begin{algorithmic}[1]
\fontsize{10pt}{10pt}\selectfont
    \Procedure{Matching}{SortedOrders}
            
    \State BuyIdx, SellIdx := 0, 0

    \State BuyOrders, SellOrders := SortedOrders

    \State MatchedOrders := []
            
    \While{BuyIdx \!\!\!\! $<$ \!\!\! len(BuyOrders) and SellIdx \!\!\! $<$ \!\!\! len(SellOrders)}

        \State BuyOrder := BuyOrders[BuyIdx]

        \State SellOrder := SellOrders[SellIdx]

        \If{(BuyOrder $\geq$ SellOrder)} %\Comment{Index in SortedOrders}
            \State MatchedOrders.\Call{Append}{(BuyOrder, SellOrder)}
            \State BuyIdx := BuyIdx + 1; SellIdx := SellIdx + 1
        \Else \State BuyIdx := BuyIdx + 1\Comment{Cannot match}
        \EndIf
    \EndWhile

    \State \Return MatchedOrders
    \EndProcedure
\end{algorithmic}
\end{algorithm}

\begin{algorithm}[H]
  \begin{algorithmic}[1]
    \caption{Bucketization - Marketplace Smart Contract}
    \label{algo:bmarket}
  \fontsize{10pt}{10pt}\selectfont
    \Do
    \State Order$_i$ := \RecvMsg{(Account$_i$, RateComm)}{Recv Order from Trader $i$}
    \State UnmatchedOrders.\Call{Append}{Order$_i$}
    \EndDo

    \item[]

    \Function{TradingRound}{UnmatchedOrders, BucketWidth}
%    \State Orderbook := [UnmatchedOrders]
    \Wait{T seconds}{Orderbook := [UnmatchedOrders]}
%    \State Orderbook.\Call{Append}{Orders}

    \State Buckets := \Call{Choose\_Buckets}{BucketWidth}

    \Comment{Choose random values}

    \State \SendMsg{Send Buckets to Traders}{Buckets}

    \Repeat
    \State \RecvMsg{(Id, Bucket$_i$, Proof$_i$)}{Recv from Trader $i$}
    \State \Call{Verify\_Range\_Proof}{Orderbook[Id], Bucket$_i$, Proof$_i$}
    \State Orderbook.\Call{Update}{Id,Bucket$_i$}
    \Until{T' seconds or all proofs received}
    
    \State SortedOrders := \Call{sort}{Orderbook, ascending}

    \State MatchedOrders := \Call{Matching}{SortedOrders}
    
    \Comment{Algorithm~\ref{algo:matching}}

    \For{(BuyOrder, SellOrder) in MatchedOrders}
    \State \RecvMsg{(Fees, Proof')}{Recv Fees from Buyer,Seller}
    \State FeesComm := BuyOrder.Comm $-$ SellOrder.Comm
    \State \Call{Verify\_Proof}{Fees, FeesComm, Proof'}
    \State \Call{Debit}{BuyOrder.Account, BuyOrder.RateComm}
    \State \Call{Credit}{SellOrder.Account, SellOrder.RateComm}
    \State \Call{Credit}{Marketplace.Account, Fees}

    \Comment{Credit to marketplace}
    \EndFor

    \State UnmatchedOrders := Orderbook $-$ MatchedOrders

    \EndFunction

  \end{algorithmic}
\end{algorithm}

\begin{algorithm}[H]
  \begin{algorithmic}[1]
    \caption{Bucketization - Trader}
    \label{algo:btrader}
  \fontsize{10pt}{10pt}\selectfont

  \Function{Trade}{Rate, Blinding, Account}

    \State RateComm := \Call{Pedersen\_Commitment}{Rate, Blinding}
    \State Id := \SendMsg{Send Order to Marketplace}{Account, RateComm}

    \State \RecvMsg{Buckets}{Recv Buckets from Marketplace}

    \State Bucket := \Call{Rate\_to\_Bucket}{Rate, Buckets}

    \State Proof := \Call{Generate\_Range\_Proof}{Rate, RateComm, Bucket}

    \State \SendMsg{Send to Marketplace}{Id, Bucket, Proof}

    \State \RecvMsg{MatchedTrader}{Recv Matching from Marketplace}

    \State $M$ := (Rate, Blinding)

    \State $\sigma$ := \Call{Digital\_Signature}{$M$}

    \State $E$ := \Call{Encrypt}{$M$, $\sigma$}

    \State $E'$ := \SendMsg{Exchange Rates with MatchedTrader}{$E$}

    \State (Rate', Blinding', $\sigma$') := \Call{Decrypt}{$E'$}

    \State \Call{Verify\_Signature}{$\sigma'$}

    \State Fees := Rate - Rate'

    \State Proof' := \Call{Generate\_Proof}{Fees, Blinding - Blinding'}

    \State \SendMsg{Send Fees to Marketplace}{Fees, Proof'}
  \EndFunction
  \end{algorithmic}
\end{algorithm}

\begin{algorithm}[H]
  \begin{algorithmic}[1]
    \fontsize{10pt}{10pt}\selectfont
    \Do
    \State Order$_i$ := \RecvMsg{(Account$_i$, ShareComms, BalProof, OrderType)}{Order from Trader $i$}
    \State Order$_i$.RateComm := \Call{Reconstruct}{ShareComms}
    \If{Order$_i$.OrderType == ``BUY''}
    \State \Call{Verify\_Range\_Proof}{Order$_i$.BalProof}
    \EndIf
    \State UnmatchedOrders.\Call{Append}{Order$_i$}
    \EndDo

    \item[]
    \Function{TradingRound}{UnmatchedOrders}
    \Wait{T seconds}{Orderbook := [UnmatchedOrders]}
%    \Wait{T seconds}
%    \State Orderbook.\Call{append}{Orders}

    \State SortedOrders := \Call{Brokers\_Sort}{Orderbook}

    \Comment{Event trigger to brokers}

    \State MatchedOrders := \Call{Matching}{SortedOrders} 

    \State Fees := \Call{Brokers\_Settle}{MatchedOrders}

    \Comment{Event trigger to brokers}

    \For{(BuyOrder, SellOrder) in MatchedOrders}
    \State \Call{Debit}{BuyOrder.Account, BuyOrder.RateComm}
    \State \Call{Credit}{SellOrder.Account, SellOrder.RateComm}
    \EndFor

    \State \Call{Credit}{Marketplace.Account, Fees}
    \State UnmatchedOrders := Orderbook $-$ MatchedOrders

    \State \Call{Brokers\_Shuffle}{ } \Comment {Event trigger to brokers}
    \EndFunction
  \end{algorithmic}
  \caption{Rialto - Marketplace Smart Contract}
  \label{algo:Mmarket}
\end{algorithm}

\vspace{-0.2in}
\begin{algorithm}[H]
  \caption{Rialto - Broker}
  \label{algo:Mbroker}
  \begin{algorithmic}[1]
    \fontsize{10pt}{10pt}\selectfont
    \Do
    \State OrderShare[Id] := \RecvMsg{(Id, RateShare, BlindingShare, Account)}{Recv Share from Trader}
    \State \Call{Verify\_Rate\_Share}{RateShare, Orders[Id]}
    \Comment{Read Order from Blockchain}
    \State \Call{Validate}{RateShare} \Comment Rialto+ (Sec~\ref{sec:validation}) 
    \State \Call{Validate}{BlindingShare} \Comment Rialto+
    \EndDo
    \item[]

    \Function{Brokers\_Sort}{Orderbook}
    \State RateShares := \IFor{OrderShare[Id].RateShare}{Id in Orderbook}
    \State \Return \Call{\hl{InvokeMPC}}{SortingMPC,RateShares}
    \EndFunction

    \item[]
    \Function {Brokers\_Settle}{MatchedOrders}
    % \State BuyerShares := \IFor{Shares[BuyerId]}{BuyerId in MatchedOrders} 
    % \State SellerShares := \IFor{Shares[SellerId]}{SellerId in MatchedOrders}
    \For{(BuyOrder, SellOrder) \textbf{in} MatchedOrders}
    \State BuyerShares.\Call{Append}{OrderShare[BuyOrder.Id].RateShare}
    \State SellerShares.\Call{Append}{OrderShare[SellOrder.Id].RateShare}
    \EndFor
    % \State (BuyerShares, SellerShares) :=

    % \IFor{(Shares[BuyOrder.Id], Shares[SellOrder.Id])}{(BuyOrder, SellOrder) \textbf{in} MatchedOrders}
    \State \Return %Fees :=
    \Call{\hl{InvokeMPC}}{SettlementMPC,(BuyerShares, SellerShares)}

    \EndFunction

    \item[]
    \Function {Brokers\_Shuffle}{}
    \State BlindingShares := \IFor{OrderShare[Id].BlindingShare}{Id in Orderbook}
    \State Accounts := \IFor{OrderShare[Id].Account}{Id in Orderbook} \Comment{Read new account balance commitments from ledger}
    \State \Return NewAccounts := \Call{\hl{InvokeMPC}}{ShuffleMPC, (BlindingShares, Accounts)}
    \EndFunction
    \end{algorithmic}
 \end{algorithm}
\vspace{-0.2in}
\begin{algorithm}[H]
  \caption{Rialto - Trader}
  \label{algo:Mtrader}
  \begin{algorithmic}[1]
    \fontsize{10pt}{10pt}\selectfont

    \Function{Trade}{Rate, Account, OrderType}
    \State RateShares := \Call{Split}{Rate, M}
    \State Blinding := \Call{Random}{ } \Comment {Acc. Bal. Re-randomzation}
    \State BlindingShares := \Call{Split}{Blinding, M}
    \State RateComms :=
    \IFor{\Call{Pedersen\_Commitment}{Share}}{Share \textbf{in} RateShares}
    \State BlindingComms :=
    \IFor{\Call{Pedersen\_Commitment}{Share}}{Share \textbf{in} BlindingShares}
    \If{OrderType == ``BUY''}
    \State BalProof := \Call{Generate\_Range\_Proof}{Rate, Account}
    \EndIf

    \State Id := \SendMsg{Send Order to Marketplace}{Account, ShareComms, BalProof, OrderType}

    \For{Broker b \textbf{in} [1,M]}
    \State \SendMsg{Send rate share to broker}{Id, \Call{Open}{RateComms[b]}, \Call{Open}{BlindingComms[b]}, Account}

    \EndFor
    \EndFunction
    \end{algorithmic}
 \end{algorithm}

%% file: paper.bbl
\begin{thebibliography}{10}
\providecommand{\url}[1]{#1}
\csname url@samestyle\endcsname
\providecommand{\newblock}{\relax}
\providecommand{\bibinfo}[2]{#2}
\providecommand{\BIBentrySTDinterwordspacing}{\spaceskip=0pt\relax}
\providecommand{\BIBentryALTinterwordstretchfactor}{4}
\providecommand{\BIBentryALTinterwordspacing}{\spaceskip=\fontdimen2\font plus
\BIBentryALTinterwordstretchfactor\fontdimen3\font minus
  \fontdimen4\font\relax}
\providecommand{\BIBforeignlanguage}[2]{{%
\expandafter\ifx\csname l@#1\endcsname\relax
\typeout{** WARNING: IEEEtran.bst: No hyphenation pattern has been}%
\typeout{** loaded for the language `#1'. Using the pattern for}%
\typeout{** the default language instead.}%
\else
\language=\csname l@#1\endcsname
\fi
#2}}
\providecommand{\BIBdecl}{\relax}
\BIBdecl

\bibitem{Forum-Report}
{Global Commodities Forum Report}, ``Trade in commodities: Challenges and
  opportunities,''
  \url{https://unctad.org/en/PublicationsLibrary/suc2015d1\_en.pdf}.

\bibitem{IOSCO-Study}
R.~Tendulkar and G.~Naacke, ``Cyber-crime, securities markets and systemic
  risk,''
  \url{https://www.iosco.org/research/pdf/swp/Cyber-Crime-Securities-Markets-and-Systemic-Risk.pdf}.

\bibitem{MtGox}
{Mt. Gox}, \url{https://en.wikipedia.org/wiki/Mt.\_Gox}.

\bibitem{Idex}
{Idex}, \url{https://idex.market/eth/idex}.

\bibitem{airswap}
{Airswap}, \url{https://www.airswap.io/}.

\bibitem{BinanceDex}
{Binance DEX}, \url{https://www.binance.org/}.

\bibitem{Bisq}
{Bisq}, \url{https://bisq.network}.

\bibitem{0xProtocol}
0x~Protocol, \url{https://0x.org}.

\bibitem{BlockDX}
{Blocknet}, \url{https://blocknet.co/block-dx/}.

\bibitem{Affogato}
{Affogato Network}, \url{https://medium.com/affogato-network}.

\bibitem{TransparentDishonesty-FC19}
S.~Eskandari, S.~Moosavi, and J.~Clark, ``Transparent dishonesty: front-running
  attacks on blockchain,'' in \emph{Financial Cryptography}, 2019.

\bibitem{FlashBoys-Arxiv19}
P.~Daian, S.~Goldfeder, T.~Kell, Y.~Li, X.~Zhao, I.~Bentov, L.~Breidenbach, and
  A.~Juels, ``Flash boys 2.0: Frontrunning, transaction reordering, and
  consensus instability in decentralized exchanges,'' 2019.

\bibitem{FrontRunning-FC19}
S.~Eskandari, S.~Moosavi, and J.~Clark, ``Sok: Transparent dishonesty:
  front-running attacks on blockchain,'' in \emph{FC}, 2019.

\bibitem{nasdaq-private-market}
{Nasdaq Private Market},
  \url{https://www.nasdaq.com/solutions/private-company-solutions}.

\bibitem{MaximalMatching-AAMAS13}
\BIBentryALTinterwordspacing
J.~Niu and S.~Parsons, ``Maximizing matching in double-sided auctions,'' in
  \emph{AAMAS}, 2013, p. 1283–1284. [Online]. Available:
  \url{http://arxiv.org/abs/1304.3135}
\BIBentrySTDinterwordspacing

\bibitem{ZeroCash}
E.~Ben-Sasson, A.~Chiesa, C.~Garman, M.~Green, I.~Miers, E.~Tromer, and
  M.~Virza, ``Zerocash: Decentralized anonymous payments from bitcoin,'' in
  \emph{IEEE SP}, 2014.

\bibitem{ZKLedger}
N.~Narula, W.~Vasquez, and M.~Virza, ``zkledger: Privacy-preserving auditing
  for distributed ledgers,'' in \emph{Usenix NSDI}, 2018.

\bibitem{Solidus}
E.~Cecchetti, F.~Zhang, Y.~Ji, A.~Kosba, A.~Juels, and E.~Shi, ``Solidus:
  Confidential distributed ledger transactions via pvorm,'' in \emph{ACM CCS},
  2017.

\bibitem{Zether}
B.~B{\"{u}}nz, S.~Agrawal, M.~Zamani, and D.~Boneh, ``Zether: Towards privacy
  in a smart contract world,'' in \emph{{FC}}, 2020, pp. 423--443.

\bibitem{Bitcoin}
S.~Nakamoto, ``Bitcoin: A peer-to-peer electronic cash system,''
  \url{http://www.bitcoin.org/bitcoin.pdf}.

\bibitem{pedersen1991threshold}
T.~P. Pedersen, ``A threshold cryptosystem without a trusted party,'' in
  \emph{Eurocrypt}, 1991, pp. 522--526.

\bibitem{MatchingAlgorithms}
K.~Janeček and M.~Kabrhel, ``Matching algorithms of international exchanges,''
  2007.

\bibitem{MatchAlgorithmsCME}
C.~Group, ``Match algorithms,''
  \url{http://web.archive.org/web/20120626161034/http://www.cmegroup.com/confluence/display/EPICSANDBOX/Match+Algorithms}.

\bibitem{ZEXE}
S.~Bowe, A.~Chiesa, M.~Green, I.~Miers, P.~Mishra, and H.~Wu, ``Zexe: Enabling
  decentralized private computation,'' in \emph{IEEE SP}, 2018, pp. 820--837.

\bibitem{TEX}
\BIBentryALTinterwordspacing
R.~Khalil, A.~Gervais, and G.~Felley, ``Tex - a securely scalable trustless
  exchange,'' 2019, \url{https://eprint.iacr.org/2019/265}. [Online].
  Available: \url{Cryptology ePrint Archive, Report 2019/265}
\BIBentrySTDinterwordspacing

\bibitem{FuturesMEX-SP18}
F.~{Massacci}, C.~N. {Ngo}, J.~{Nie}, D.~{Venturi}, and J.~{Williams},
  ``Futuresmex: Secure, distributed futures market exchange,'' in \emph{IEEE
  SP}, 2018, pp. 335--353.

\bibitem{darkMPC}
J.~Cartlidge, N.~P. Smart, and Y.~Talibi~Alaoui, ``Mpc joins the dark side,''
  in \emph{ACM AsiaCCS}, 2019, pp. 148--159.

\bibitem{P2DEX}
C.~Baum, B.~David, and T.~K. Frederiksen, ``P2dex: Privacy-preserving
  decentralized cryptocurrency exchange,'' in \emph{Applied Cryptography and
  Network Security}, 2021, pp. 163--194.

\bibitem{cormen2001section}
T.~Cormen, C.~Leiserson, R.~Rivest, and C.~Stein, ``Section 8.4: Bucket sort,''
  \emph{Introduction to Algorithms,}, pp. 174--177, 2001.

\bibitem{waksman}
A.~Waksman, ``A permutation network,'' \emph{Journal of the ACM (JACM)},
  vol.~15, no.~1, pp. 159--163, 1968.

\bibitem{schnorr}
C.-P. Schnorr, ``Efficient identification and signatures for smart cards,'' in
  \emph{Crypto}, 1989, pp. 239--252.

\bibitem{bunz2018bulletproofs}
B.~B{\"u}nz, J.~Bootle, D.~Boneh, A.~Poelstra, P.~Wuille, and G.~Maxwell,
  ``Bulletproofs: Short proofs for confidential transactions and more,'' in
  \emph{IEEE SP}, 2018, pp. 315--334.

\bibitem{secretsharing}
A.~Shamir, ``How to share a secret,'' \emph{Commun. ACM}, vol.~22, no.~11, p.
  612–613, 1979.

\bibitem{10.1007/3-540-45749-6_34}
K.~Deshmukh, A.~V. Goldberg, J.~D. Hartline, and A.~R. Karlin, ``Truthful and
  competitive double auctions,'' in \emph{ESA}, 2002, pp. 361--373.

\bibitem{wang2016truthful}
S.~Wang, ``Truthful double auction mechanisms for heterogeneous spectrums and
  spectrum group-buying,'' Ph.D. dissertation, 2016.

\bibitem{abe2001remarks}
M.~Abe and F.~Hoshino, ``Remarks on mix-network based on permutation
  networks,'' in \emph{International Workshop on Public Key Cryptography},
  2001, pp. 317--324.

\bibitem{ron2013quantitative}
D.~Ron and A.~Shamir, ``Quantitative analysis of the full bitcoin transaction
  graph,'' in \emph{FC}, 2013, pp. 6--24.

\bibitem{reid2013analysis}
F.~Reid and M.~Harrigan, ``An analysis of anonymity in the bitcoin system,'' in
  \emph{Security and privacy in social networks}.\hskip 1em plus 0.5em minus
  0.4em\relax Springer, 2013, pp. 197--223.

\bibitem{Fabric-Eurosys19}
E.~Androulaki, A.~Barger, V.~Bortnikov, C.~Cachin, K.~Christidis, A.~De~Caro,
  D.~Enyeart, C.~Ferris, G.~Laventman, and Y.~Manevich, ``Hyperledger fabric: A
  distributed operating system for permissioned blockchains,'' in \emph{ACM
  Eurosys}, 2018.

\bibitem{MPSPDZ}
C.~Data61, ``Mp-spdz,'' \url{https://github.com/data61/MP-SPDZ}.

\bibitem{keller2020mp}
M.~Keller, ``Mp-spdz: A versatile framework for multi-party computation,'' in
  \emph{Proceedings of the 2020 ACM SIGSAC Conference on Computer and
  Communications Security}, 2020, pp. 1575--1590.

\bibitem{Bulletproofs}
D.~Cryptography, \url{https://github.com/dalek-cryptography/bulletproofs}.

\bibitem{amihud1986asset}
Y.~Amihud, H.~Mendelson \emph{et~al.}, ``Asset pricing and the bid-ask
  spread,'' \emph{Journal of financial Economics}, vol.~17, no.~2, pp.
  223--249, 1986.

\bibitem{kraft2019game}
D.~Kraft, ``Game-theoretic randomness for blockchain games,'' \emph{arXiv
  preprint arXiv:1901.06285}, 2019.

\bibitem{algorand}
Y.~Gilad, R.~Hemo, S.~Micali, G.~Vlachos, and N.~Zeldovich, ``Algorand: Scaling
  byzantine agreements for cryptocurrencies,'' in \emph{SOSP}, 2017, pp.
  51--68.

\bibitem{praos}
B.~David, P.~Ga{\v{z}}i, A.~Kiayias, and A.~Russell, ``Ouroboros praos: An
  adaptively-secure, semi-synchronous proof-of-stake blockchain,'' in
  \emph{Eurocrypt}, 2018, pp. 66--98.

\bibitem{Blom}
G.~Blom, ``Statistical estimates and transformed beta-variables,'' Ph.D.
  dissertation, Almqvist \& Wiksell, 1958.

\bibitem{berg2009hanson}
H.~Berg, T.~A. Proebsting \emph{et~al.}, ``Hanson's automated market maker,''
  \emph{Journal of Prediction Markets}, vol.~3, no.~1, pp. 45--59, 2009.

\bibitem{xu2021sok}
J.~Xu, N.~Vavryk, K.~Paruch, and S.~Cousaert, ``Sok: Decentralized exchanges
  (dex) with automated market maker (amm) protocols,'' \emph{arXiv preprint
  arXiv:2103.12732}, 2021.

\bibitem{capponi2021adoption}
A.~Capponi and R.~Jia, ``The adoption of blockchain-based decentralized
  exchanges: A market microstructure analysis of the automated market maker,''
  \emph{Available at SSRN 3805095}, 2021.

\end{thebibliography}
